# Accurate estimation of dynamical quantities for nonequilibrium nanoscale system

Zhi Xu[1,2], Han Li[1,2], Ming Ma[1,2]*

[1]Department of Mechanical Engineering, State Key Laboratory of Tribology, Tsinghua University, Beijing 100084, China

[2]Center for Nano and Micro Mechanics, Tsinghua University, Beijing 100084, China

*Email: maming16@tsinghua.edu.cn

# Abstract

Fluctuations of dynamical quantities are fundamental and inevitable. For the booming research in nanotechnology, huge relative fluctuation comes with the reduction of system size, leading to large uncertainty for the estimates of dynamical quantities. Thus, increasing statistical efficiency, i.e., reducing the number of samples required to achieve a given accuracy, is of great significance for accurate estimation. Here we propose a theory as a fundamental solution for such problem by constructing auxiliary path for each real path. The states on auxiliary paths constitute canonical ensemble and share the same macroscopic properties with the initial states of the real path. By implementing the theory in molecular dynamics simulations, we obtain a nanoscale Couette flow field with an accuracy of 0.2 μm/s with relative standard error < 0.1. The required number of samples is reduced by 12 orders compared to conventional method. The predicted thermolubric behavior of water sliding on a self-assembled surface is directly validated by experiment under the same velocity. As the theory only assumes the system is initially in thermal equilibrium then driven from that equilibrium by an external perturbation, we believe it could serve as a general approach for extracting the accurate estimate of dynamical quantities from large fluctuations to provide insights on atomic level under experimental conditions, and benefit the studies on mass transport across (biological) nanochannels and fluid film lubrication of nanometer thickness.

# Introduction

During the past decades, the scale of the systems people are interested in keeps on decreasing, from micro, nano to angstrom, with typical applications like water desalination, nanopore sequencing, and heat dissipation of the chip (*1-4*). This trend leads to the number of particles constituting the system, $N$, reducing significantly as it is proportional to the cubic of the scale. It also leads to large relative fluctuation (RF) of the dynamical quantities of the systems which is inverse proportional to the square root of $N$ (*5*). As a result, the standard error (SE) of the mean estimate for dynamical quantity $A$ with multiple measurements ($\bar{A}$) is large. In other words, the confidence intervals $[\bar{A} - \text{SE}, \bar{A} + \text{SE}]$ from numerous samples that encompass the exact value of $A$ with a probability of 68% are broad, leading to large uncertainties in the measurement. Thus, the intrinsic large RF brings difficulty to the estimate of the dynamical quantity for both experiments (*6, 7*) and simulations (*8-10*) with nanoscale systems.

The conventional method uses the path ensemble average to estimate the quantity at nonequilibrium. The accuracy of the estimate is limited by finite sampling. To increase the accuracy of the estimate, several strategies have been proposed. First, since SE is inverse proportional to the square root of the number of independent samples ($n$), i.e., $\text{SE} \propto 1/\sqrt{n}$, some methods increase $n$ to get a good estimation, like using parallel replica dynamics with rare events (*11*), reducing the cost of single sampling (*12*). Second, some studies lead to a practical closed form solution for the uncertainty of the estimation (*13, 14*) and present new sampling algorithms to increase the statistical efficiency by one to three orders of magnitude. Third, as the statistical uncertainty of an estimator is usually quantified by its SE in the asymptotic limit, some studies find an optimal estimator (*15, 16*) which reduces this asymptotic SE. These studies could increase the statistical efficiency by three orders.

With these methods, however, for nanoscale system with large RF, it still requires a large quantity of sampling to provide enough data to get accurate estimate, e.g., with relative standard error (RSE) < 0.1 (*9, 17*). As to water, Fig. 1a shows the RF of the centre of mass velocity ($v_{\text{com}}$) varies with the number of particles $N$ at room temperature ($T$ = 298 K). For a steady water flow, RF = $\sigma_{v_{\text{com}}}/v_{\text{com}}$ where $\sigma_{v_{\text{com}}} = \sqrt{1/\beta N m_{\text{water}}}$, $\beta = 1/k_B T$ and $m_{\text{water}}$ is the mass of individual water molecule. Derivation of $\sigma_{v_{\text{com}}}$ is given in Supplementary information (SI) Section 1. With $v_{\text{com}} = 10^{-4}$ m/s which is a typical value for experiments (17), it is clear that RF at nano or angstrom scale is huge (>$10^4$). For water flow at nanoscale (e.g., $N$ = 10,000), Fig. 1b illustrates that with the first kind of method, how many independent samples ($n$) are required to detect certain values of $v_{\text{com}}$ with a given RSE at $T$=298 K. Here RSE = $\sigma_{v_{\text{com}}}/\sqrt{n}v_{\text{com}}$ is used to measure the reliability of the estimate as the confidence interval is $[1 - \text{RSE}, 1 + \text{RSE}]v_{com}$, i.e., small RSE leading to accurate estimation. The number of samples needed to give an accurate estimate of $v_{\text{com}} = 10^{-4}$ m/s is 5 to 7 orders larger than that in the existing research (9, 18-22). For the second and third kinds of methods as mentioned above, though they can increase the statistical efficiency up to 3 orders, the required number of samples would still be 2 - 4 orders larger than existing studies, far beyond the current ability of sampling. Therefore, for water at nanoscale, the existing methods to get accurate estimate of $v_{\text{com}}$ comparable to experiments (e.g., $10^{-4}$ m/s) are inefficient and become even impractical.

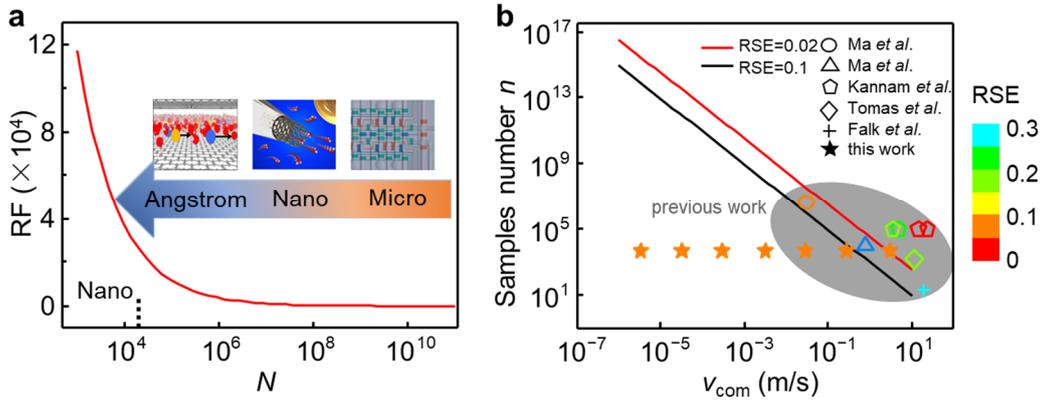

**Figure 1. The number of samples needed to estimate the flow rate of the nanoconfined water. a,** Relative fluctuation (RF) as a function of the number of particles $N$ in the system. The arrow shows the recent experimental systems from micro (23), nano (24) to angstrom (25) scale, and the dashed line indicates the number of water molecules in a cube with a side length of 100 nm. **b,** The number of independent

samples *n* needed to estimate the velocity $v_{com}$ with given relative standard error (RSE). The values of $v_{com}$ for nanoconfined water in existing molecular simulations (0.03 to 24.3 m/s) and this work ($3.36\times10^{-6}$ to 3 m/s) are plotted as different symbols.

In this paper, we propose a fundamentally different method to efficiently reduce the standard error of the estimate for dynamical quantities at nonequilibrium. Through generating an auxiliary path corresponding to each path in the phase space with a specific protocol, the path ensemble average of the difference between the quantities in real path and auxiliary path is found to be a good estimate of the quantities. This method is thus called auxiliary path method (APM). For achieving the same accuracy, the number of samples required for APM is much smaller than the conventional method.

## Auxiliary path method

For a classical system composed of *N* particles, a dynamical quantity *A* depends upon time *t* via the time dependence of the coordinates $\mathbf{r}_1(t), ..., \mathbf{r}_N(t)$ and momenta $\mathbf{p}_1(t), ..., \mathbf{p}_N(t)$ of the particles, i.e., $A(t) = A[\mathbf{r}_1(t), ..., \mathbf{r}_N(t), \mathbf{p}_1(t), ..., \mathbf{p}_N(t)]$. For each path, the phase space point $x_t \equiv [\mathbf{r}_1(t), ..., \mathbf{r}_N(t), \mathbf{p}_1(t), ..., \mathbf{p}_N(t)]$ is determined by integrating Newton's laws from the initial microstate (*t* = 0) with the corresponding phase space point $x_0 \equiv [\mathbf{r}_1(0), ..., \mathbf{r}_N(0), \mathbf{p}_1(0), ..., \mathbf{p}_N(0)]$, i.e. the dynamics of the system is deterministic. The exact value of *A* at time *t*, $\langle A \rangle_t$, can be defined via path ensemble average as $\langle A \rangle_t = \int dx_0 F(x_0) A(t)$, where $F(x_0)$ is the distribution function of initial microstates in phase space (*26*). In this paper, the path ensemble is constituted by the initial thermal equilibrium and the process by which the system is subsequently perturbed from that equilibrium as suggested by Crooks (*27*). For system initially corresponds to canonical ensemble, $F(x_0) = e^{-\beta H(x_0)}/\int dx_0 e^{-\beta H(x_0)}, \beta = 1/k_B T$ and $H(x_0)$ is the Hamiltonian of the initial microstate. For *Q* initial microstates correspond to a given sampling of the canonical phase space distribution of a system, upon perturbation, they will generate *Q* nonequilibrium paths as time propagates. Thus we have $\langle A \rangle_t = \lim_{Q \to \infty} \frac{1}{Q} \sum_{i=1}^{Q} A_i(t)$ (*28*), where *i* indicates the *i*th path.

While $\langle A \rangle_t$ is the value people are interested in, practical sampling hardly fulfills ergodicity in the path ensemble. Therefore, the conventional method usually uses finite $m$ paths in the path ensemble to estimate $\langle A \rangle_t$ as $\bar{A}_m(t) = \frac{1}{m}\sum_{i=1}^{m} A_i(t)$, with $\langle A \rangle_t = \lim_{m \to \infty} \bar{A}_m(t)$. According to central limit theorem, the standard error of $\bar{A}_m(t)$ can be estimated as $\sigma_A/\sqrt{m}$, where $\sigma_A$ is the standard deviation of quantity $A$.

Now assuming that for the same system we have auxiliary path corresponding to each path in the nonequilibrium path ensemble, and the states $x'_t \equiv [\mathbf{r}'_1(t), \ldots, \mathbf{r}'_N(t), \mathbf{p}'_1(t), \ldots, \mathbf{p}'_N(t)]$ on all auxiliary paths are required to constitute canonical ensemble at any time $t$ when $Q$ goes to infinity. Now we define a quantity $B(t) = A(t) - A'(t)$, here $A'(t) = A(x'_t)$. The path ensemble average $\langle B \rangle_t = \lim_{Q \to \infty} \frac{1}{Q} \sum_{i=1}^{Q}(A_i(t) - A'_i(t)) = \lim_{Q \to \infty} \frac{1}{Q} \sum_{i=1}^{Q} A_i(t) - \lim_{Q \to \infty} \frac{1}{Q} \sum_{i=1}^{Q} A'_i(t)$. For the former, $\lim_{Q \to \infty} \frac{1}{Q} \sum_{i=1}^{Q} A_i(t) = \langle A \rangle_t$. For the latter, according to the assumption that the auxiliary states $x'_t$ constitute canonical ensemble at any time $t$, we have $\lim_{Q \to \infty} \frac{1}{Q} \sum_{i}^{Q} A'_i(t) = \langle A \rangle_{EC}$, where $\langle A \rangle_{EC}$ is the canonical ensemble average of $A$. Thus, $\langle B \rangle_t = \langle A \rangle_t - \langle A \rangle_{EC}$, i.e.

$$\langle A \rangle_t = \langle B \rangle_t + \langle A \rangle_{EC}. \qquad (1)$$

This is the proposed calculation method of $\langle A \rangle_t$, and it's the core of APM. Its efficiency over the conventional method which estimates $\langle A \rangle_t$ as $\bar{A}_m(t) = \frac{1}{m}\sum_{i=1}^{m} A_i(t)$ can be quantified via dimensionless quantity $\gamma$ as

$$\text{SE}_{\text{APM}} = \gamma \text{SE}_{\text{CON}}, \qquad (2)$$

where $\text{SE}_{\text{APM}}$ and $\text{SE}_{\text{CON}}$ are the standard error of APM and the conventional method respectively. To achieve the same SE, the APM reduces the required number of samples by $1/\gamma^2$ times compared to the conventional method due to $\text{SE} \propto 1/\sqrt{n}$. In other words, the statistical efficiency of APM is increased by $1/\gamma^2 - 1$ times compared to that of conventional method. On the basis that the auxiliary paths satisfy the requirements, according to equation (1), $\langle A \rangle_t$ can be estimated by the sum of $\bar{B}_m(t) = \frac{1}{m}\sum_{i=1}^{m} B_i(t)$ and a known constant $\langle A \rangle_{EC}$. Now $\text{SE}_{\text{APM}} = \sigma_B/\sqrt{m}$ where $\sigma_B$ is the standard deviation of quantity B and $\text{SE}_{\text{CON}} = \sigma_A/\sqrt{m}$, thus $\gamma = \sigma_B/\sigma_A$. The quantity $\sigma_B$ can be estimated as $\sqrt{\sum_{i=1}^{m}(B_i(t) - \bar{B}_m(t))^2/m}$ and $\sigma_A$ can be estimated as $\sqrt{\sum_{i=1}^{m}(A_i(t) - \bar{A}_m(t))^2/m}$, resulting in $\gamma = \sqrt{\frac{\overline{B^2}_m(t) - (\bar{B}_m(t))^2}{\overline{A^2}_m(t) - (\bar{A}_m(t))^2}}$, where $\overline{B^2}_m(t) = \frac{1}{m}\sum_{i=1}^{m}B_i^2(t)$ and $\overline{A^2}_m(t) = \frac{1}{m}\sum_{i=1}^{m}A_i^2(t)$. For systems with large relative fluctuation, $\overline{A^2}_m(t) \gg (\bar{A}_m(t))^2$, thus $\gamma \approx \sqrt{\frac{\overline{B^2}_m(t) - (\bar{B}_m(t))^2}{\overline{A^2}_m(t)}} < \sqrt{\overline{B^2}_m(t)/\overline{A^2}_m(t)}$. When $B^2(t)$ is much smaller than $A^2(t)$, $\gamma \ll 1$, which means $\text{SE}_{\text{APM}}$ is much smaller than $\text{SE}_{\text{CON}}$. Thus, the auxiliary path method, i.e., using $\bar{B}_m(t) + \langle A \rangle_{\text{EC}}$ to estimate $\langle A \rangle_t$, will be much more accurate than the conventional method, i.e., using $\bar{A}_m(t)$ to estimate $\langle A \rangle_t$.

Generally speaking, the APM is done by designing auxiliary path corresponding to each real path in the nonequilibrium path ensemble. The requirements for the auxiliary paths are that the auxiliary states $x_t'$ constitute canonical ensemble and share the same set of macroscopic properties, e.g., number and types of particles, volume, temperature, with the initial canonical ensemble at any time $t$. Based on both requirements, the closer the states on auxiliary path and real path are, the smaller $\gamma$ is, the more effective this method is.

## A specific protocol for APM

Next, we present a specific protocol to generate auxiliary path as shown in Fig. 2a. Notice that the dynamics of the system remains unchanged, only the extra auxiliary paths are constructed to increase the accuracy of the estimate. We limit the studied systems as those which can preserve the canonical ensemble if having not been perturbed. At any given time $t$, there are three kinds of states, which are the real states $x_t = [\mathbf{r}_1(t), \ldots, \mathbf{r}_N(t), \mathbf{p}_1(t), \ldots, \mathbf{p}_N(t)]$, the states $x_t' = [\mathbf{r}_1'(t), \ldots, \mathbf{r}_N'(t), \mathbf{p}_1'(t), \ldots, \mathbf{p}_N'(t)]$ and the middle states $x_t''$ which will be illustrated below. The basic steps of the protocol can be summarized as follows:

1. At $t = 0$, the three types of states, $x_t$, $x_t'$ and $x_t''$, are exactly the same and constitute canonical ensemble.

2. Integrating the equations of motion forward in time under perturbation with a timestep of $\delta t$ to generate the real states $x_{t+\delta t}$ from $x_t$.
3. Integrating the equations of motion forward in time without perturbation to generate the states $x'_{t+\delta t}$ from $x''_t$. Note that at $t = 0$, $x''_t$ and $x_t$ are the same.
4. The middle states $x''_{t+\delta t}$ are obtained by combining the coordinate $[\mathbf{r}_1(t + \delta t), ..., \mathbf{r}_N(t + \delta t)]$ and the momentum $[\mathbf{p}'_1(t + \delta t), ..., \mathbf{p}'_N(t + \delta t)]$, i.e. $[\mathbf{r}_1(t + \delta t), ..., \mathbf{r}_N(t + \delta t), \mathbf{p}'_1(t + \delta t), ..., \mathbf{p}'_N(t + \delta t)]$.
5. Update $t$ to $t + \delta t$ and repeat steps 2 to 4.

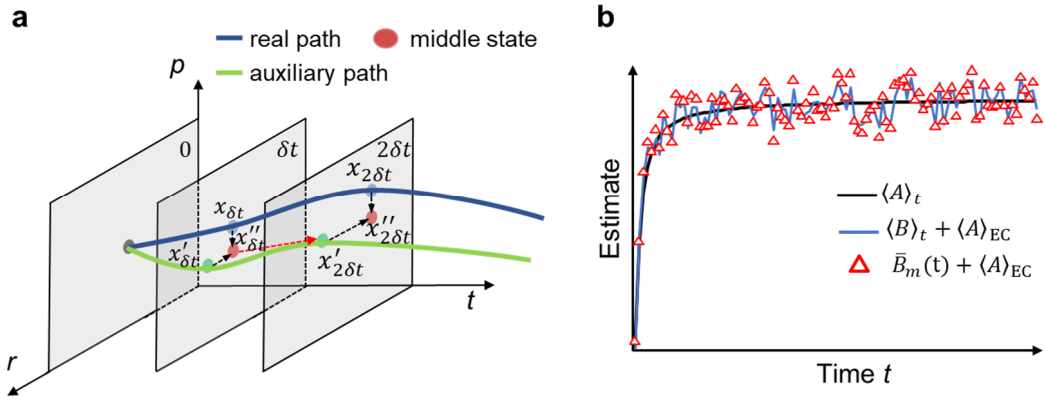

**Figure 2**. **Schematic of a specific protocol for APM. a,** The three axes are coordinate $r$, momentum $p$, and time $t$. The blue solid line stands for a real path in the nonequilibrium path ensemble with external perturbation and the green solid line is the corresponding auxiliary path. The black arrows reveal the origin of the middle states and the red arrow represents a relaxation with a duration of $\delta t$. **b,** A schematic of the error analysis for the specific protocol. Two parts of error are introduced. One is the bias of $\langle B \rangle_t + \langle A \rangle_{EC}$ (blue curve) from $\langle A \rangle_t$ (black curve) caused by the finite $\lambda$. The other is the error caused by the estimation of $\langle B \rangle_t + \langle A \rangle_{EC}$ (blue curve) with $\bar{B}_m(t) + \langle A \rangle_{EC}$ (red triangles).

Then we will show that the states $x'_t$ fulfill the requirement for auxiliary states approximately, i.e., they constitute canonical ensemble and share the same set of macroscopic properties with the initial canonical ensemble. Naturally at the beginning, both the states $x'_0$ and $x''_0$ constitute canonical ensembles as they are identical to $x_0$.

For states $x'_{\delta t}$, as they are propagated from $x''_0$ without any perturbation, they also constitute canonical ensemble, in other words, the probability of states $x'_{\delta t}$, $f(x'_{\delta t}) \propto e^{-\beta H(x'_{\delta t})}$. Further, the essence of this protocol is a one-to-one correspondence between the real path and the auxiliary path, hence each middle state $x''_{\delta t}$ is determined by the auxiliary state $x'_{\delta t}$ and its unique corresponding real state, which means $f(x''_{\delta t}) = f(x'_{\delta t})$. When the condition $\lambda(\delta t) \to 0$ is met, where $\lambda(\delta t) = \beta \left( \max_i \{\Delta H_i(\delta t)\} - \min_i \{\Delta H_i(\delta t)\} \right)$, $\Delta H_i(\delta t) = H_i(x'_{\delta t}) - H_i(x''_{\delta t})$, we can derive that $f(x''_{\delta t}) = \frac{e^{-\beta H(x''_{\delta t})}}{\sum_i e^{-\beta H(x''_{\delta t})}}$ (full derivation is given in SI Section 2). It means the states $x''_{\delta t}$ satisfy canonical ensemble distribution function, thus constitute canonical ensemble. The auxiliary states $x'_{2\delta t}$ are obtained by a relaxation time of $\delta t$ from the middle states $x''_{\delta t}$, hence, preserve the canonical ensemble. At this point, the construction implemented the auxiliary states from one moment $(x'_{\delta t})$ to the next $(x'_{2\delta t})$, as the construction goes on, we can get the auxiliary states $x'_t$ at any time $t$ when $\lambda(t) \to 0$.

With such protocol based on APM, $\langle A \rangle_t$ can be estimated as $\bar{B}_m(t) + \langle A \rangle_{EC}$. As shown in Fig. 2b, two parts of error are introduced. One is the bias of $\langle B \rangle_t + \langle A \rangle_{EC}$ from $\langle A \rangle_t$ caused by the finite $\lambda$ which characterizes the approximation in the construction process of the auxiliary path. The other is the error caused by the estimation of $\langle B \rangle_t + \langle A \rangle_{EC}$ with $\bar{B}_m(t) + \langle A \rangle_{EC}$ due to the finite sampling. For the bias, when $\lambda = 0$, equation (1) shows $\langle A \rangle_t = \langle B \rangle_t + \langle A \rangle_{EC}$. However, the finite small $\lambda$ will cause a bias of the auxiliary states away from the canonical ensemble. As a result, the upper limit of the bias is $\sigma_A \sum_{j=1}^{C-1} \alpha^{C-j} \lambda(j\delta t)$, where $\alpha = e^{-\delta t/\tau}$, $C$ is the number of timesteps accumulated, $t = C\delta t$ and $\tau$ is the relaxation time of the quantity $A$ for a given system (full derivation is given in SI Section 3). For the error caused by the estimation of $\langle B \rangle_t + \langle A \rangle_{EC}$ with $\bar{B}_m(t) + \langle A \rangle_{EC}$, the confidence interval is $\left[ \bar{B}_m(t) + \langle A \rangle_{EC} - \frac{\sigma_B}{\sqrt{m}}, \bar{B}_m(t) + \langle A \rangle_{EC} + \frac{\sigma_B}{\sqrt{m}} \right]$ where $\frac{\sigma_B}{\sqrt{m}}$ is the standard error.

Therefore, for any instantaneous moment, $\bar{B}_m(t) + \langle A \rangle_{EC}$ is a biased estimator of $\langle A \rangle_t$ with the upper limit of the bias being $\sigma_A \sum_{j=1}^{C-1} \alpha^{C-j} \lambda(j\delta t)$. During the steady state where $\langle A \rangle_t$ is a constant, however, the average of $\bar{B}_m(t) + \langle A \rangle_{EC}$ calculated with $C_s$ timesteps $\frac{1}{C_s}\sum_{C_s} \bar{B}_m(t) + \langle A \rangle_{EC}$ is an unbiased estimator due to the randomness of the bias between $\langle B \rangle_t + \langle A \rangle_{EC}$ and $\langle A \rangle_t$ (see details in SI Section 3). According to central limit theorem, it has a standard error of $\frac{1}{\sqrt{C_s}}\left(\frac{\sigma_B}{\sqrt{m}} + \sigma_A \sum_{j=1}^{N-1} \alpha^{N-j} \lambda(j\delta t)\right)$. With the same number of samples, the standard error for the conventional method of using $\frac{1}{C_s}\sum_{C_s} \bar{A}_m(t)$ to estimate $\langle A \rangle_t$ is $SE_{con} = \sigma_A/\sqrt{mC_s}$. Therefore, $\gamma = \sigma_B/\sigma_A + \sqrt{m}\sum_{j=1}^{C-1} \alpha^{C-j}\lambda(j\delta t)$, and the times of increasement in the statistical efficiency of this specific protocol for APM over conventional method can be estimated as $\frac{1}{\gamma^2} - 1$.

## Typical applications

The auxiliary path method brings us a powerful theoretical tool to obtain accurate estimate of dynamical quantities for nanoscale systems with intrinsically large relative fluctuations. This is particularly important for solid - liquid interface at nanoscale, as at room temperature the thermal fluctuation for liquid next to the solid surface is very large compared to the dynamical quantities people are interested in, e.g., flow rate. Among the many topics for solid – liquid interface, nanofluidics (*29*) and liquid superlubricity (*30*) have attracted great attentions due to their great applications in energy and friction reduction. Therefore, we will take both cases as typical applications to demonstrate the ability of APM in increasing the statistical efficiency.

We first focused on a typical yet simple system where water is sheared by two parallel double-layer graphene sheets (homogeneous surfaces). Such a system (Fig. 3a) is a typical Couette flow, and it has profound significance in the field of nanofluidics. The length and width of the graphene layer are about 5 nm and 3 nm, respectively. The height of the channel is about 2.5 nm. The velocity of each water molecule is the dynamical quantity we are interested in. In the simulations, periodic boundary conditions were applied along the *x* and *y* directions parallel to the sheets, and a spring with stiffness of 2.7 N/m along *z* direction was applied on each carbon atom in the top and bottom layers of graphene, connecting their present positions to their equilibrium positions. The water molecules were sheared by the graphene sheets moving at a constant velocity ($V_{wall}$) in opposite directions along *x*-axis (Fig. 3a). All the simulations were performed using LAMMPS (*31*) with a self-developed package incorporating the protocol for APM proposed above. A full description of the molecular dynamics (MD) simulation procedure is given in SI Section 4.

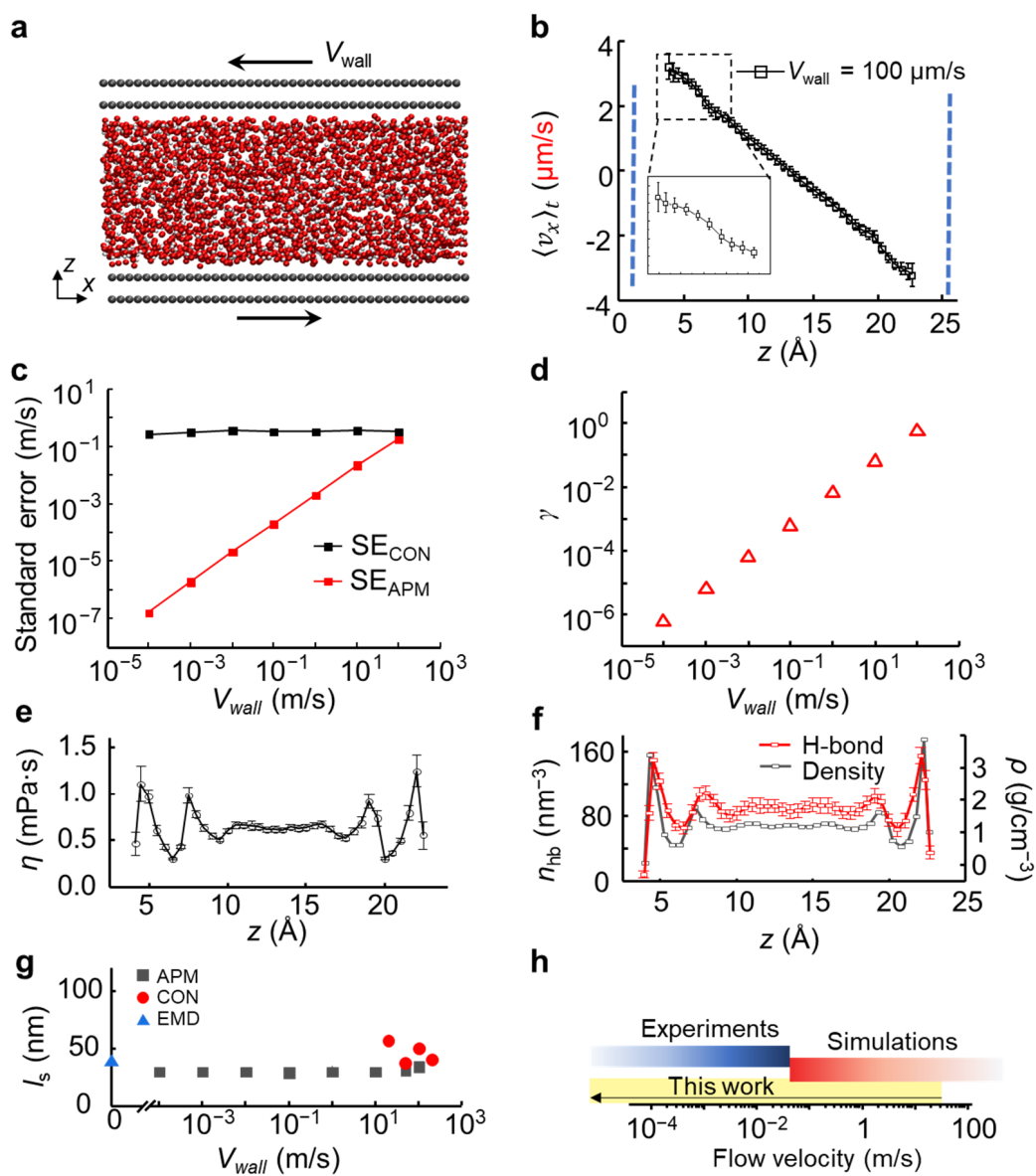

**Figure 3. Application of the specific protocol to nanoconfined water flow with homogenous surfaces. a,** Side view of a typical studied system for water sheared by two double-layer graphene sheets. **b,** The velocity profile of water flow obtained from APM with $V_{\text{wall}} = 100$ μm/s. The blue dotted lines are the average position of the innermost walls of graphene sheets. **c,** Comparison between the standard error of velocity profiles obtained by the conventional method and APM with the same number of samples. **d,** The ratio $\gamma$ defined as $SE_{\text{APM}}/SE_{\text{CON}}$. **e,** The local viscosity $\eta$ of water at different distance away from the innermost walls of graphene sheets. **f,** The density of water $\rho$ and the number density of hydrogen bond $n_{\text{hb}}$ at different distance away from the innermost walls of graphene sheets. **g,** The slip length ($l_s$) versus $V_{\text{wall}}$. The *x*-axis is composed of linear and logarithmic parts, separated by the break line. The only deviation is $l_s$ based on the conventional method at $V_{\text{wall}} = 20$ m/s, where the velocity profile is covered by the thermal noise. **h,** Comparison of the flow velocity ranges studied in previous experiments ($1\times10^{-6}$ to $4\times10^{-2}$ m/s) (*9*), previous simulations ($3\times10^{-2}$ to 1000 m /s) (*9, 18-22*), and the present work ($3.36\times10^{-6}$ to 3 m/s).

According to equation (1) for APM, the velocity field of water $\langle \mathbf{v} \rangle_t = \lim_{Q\to\infty} \frac{1}{Q} \sum_{i=1}^{Q} \left( \mathbf{v}_i(t) - \mathbf{v}'_i(t) \right) + \langle \mathbf{v} \rangle_{\text{EC}}$ where $\mathbf{v}_i(t)$ is the velocity field of state $x_t$ on the *i*th real path and $\mathbf{v}'_i(t)$ is the velocity field of state $x'_t$ on the corresponding auxiliary path. The equilibrium ensemble average of velocity field $\langle \mathbf{v} \rangle_{\text{EC}} = 0$. The perturbation to the water is the shear imposed by the moving graphene layers. With the same macroscopic properties, 10 independent simulations were carried at a given $V_{\text{wall}}$. During steady state, for each simulation, sampling over $\mathbf{v}_i(t)$ and $\mathbf{v}'_i(t)$ was conducted every 1 ps, lasting for 500 ps. These amount to a total number of samples of 5000 ($m = 10$, $C_s = 500$). Details of data analysis can be found in SI Section 5.

The profile along *z* of the flow velocity in the *x* direction ($\langle v_x \rangle_t$) during steady state is shown in Fig. 3b. It is worth noting that for $V_{\text{wall}} = 100$ μm/s, an accuracy of 0.2 μm/s with RSE < 0.1 was obtained for $\langle v_x \rangle_t$. To the best of our knowledge, this is the lowest velocity reported for water flow examined by MD, four orders smaller than existing studies (> 0.03 m/s) (*9*). To understand why we could obtain such a low flow velocity with high accuracy, we calculated the corresponding $SE_{\text{CON}}$ and $SE_{\text{APM}}$ as $\sigma_{v_x}/\sqrt{mC_s}$ and $\frac{1}{\sqrt{C_s}}\left(\frac{\sigma_{v_x-v'_x}}{\sqrt{m}} + \sigma_{v_x}\sum_{j=1}^{C-1}\alpha^{C-j}\lambda(j\delta t)\right)$ respectively. The correspondence to equation (1) $\langle A \rangle_t = \langle B \rangle_t + \langle A \rangle_{EC}$ is $v_x$ being $A$ and $v_x - v'_x$ being $B$. The values of $\sigma_{v_x}$ and $\sigma_{v_x-v'_x}$ were obtained from the simulations, with $m = 10$, $C_s = 500$ and $\alpha = e^{-t/\tau}$ where $t = 1$ fs and $\tau = 0.1$ ps for the velocity relaxation time of water molecule with SPC/E force field (*32*). From Fig. 3c, it is evident that SE$_{APM}$ is much smaller than SE$_{CON}$ for $V_{wall} <$ 10 m/s. For $\gamma$ = SE$_{APM}$/SE$_{CON}$, it reduces as $V_{wall}$ decreases (Fig. 3d). It is worth noting that with $V_{wall}$ = 100 μm/s, $\gamma = 6.1 \times 10^{-7}$. In other words, the APM reduces the required number of samples by $2.2 \times 10^{12}$ ($1/\gamma^2$) times compared to the conventional method, i.e., achieves a 12 orders increasement in statistical efficiency. It is worth noting that to obtain a velocity profile as shown in Fig. 3b, the conventional method would take 74 billion years with a CPU core running at 2.9 GHz, which is far beyond the present computing power (see SI Section 8 for more details of the estimation).

Such a high accuracy provides us an unprecedented opportunity to examine the fine structure of the flow field at nanoscale. The linear profile (Fig. 3b), away from the boundary region (0.7 nm away from the graphene specifically (*18, 33*)), is met with the classical prediction of planar Couette flow. However, we found a slight but clear oscillation of $\langle v_x \rangle_t$ for water close to the surface (the inset of Fig. 3b) thanks to the high accuracy of APM. The oscillation of velocity at interface has been found in simulations of simple fluid with a large shear rate ($10^{11}$ s$^{-1}$) (*34, 35*). For water, the shear rate studied here is $4 \times 10^4$ s$^{-1}$, which is matched with the shear rate used in experiments ($10^4$ s$^{-1}$ to $3 \times 10^5$ s$^{-1}$) (*36, 37*).

In Fig. 3e, we plotted the local viscosity $\eta(z) = \tau / \frac{d\langle v_x \rangle_t}{dz}$, where $\tau$ is the shear stress that holds constant in the stable flows. Previous studies (*18, 33*) have found that the viscosity of water at the interface deviates from that of the bulk phase, but the detailed distribution remains unknown. The theory (*34, 38*) for the viscosity distribution of confined Lennard-Jones fluid has been established using the local average density model (*39*). Thanks to the accurate velocity profile from the APM (Fig. 3b), we achieved the quantitative calculation of viscosity distribution for nanoconfined water for the first time (see the calculation details in SI Section 6). The oscillation of viscosity in the boundary region has a period of 0.25 nm, consistent with that of the density for water (Fig. 3f). The viscosity of nanoconfined water has a 4 times variation, ranging from 0.30 to 1.25 mPa·s. This is different from the results of applying local average density model with water (*38*). Such difference shows that the local average density model may be not suitable for confined water of which the viscosity is not only affected by density but also by structure (e.g., the distribution of hydrogen bonds as shown in Fig. 3f) (*40*).

The accurate velocity profile (Fig. 3b) also enables us to calculate the slip length ($l_s$) which is a typical quantity characterizing the flow profile and resistance at the liquid-solid interface (*41*). The method in Ref. (*42*) for calculating $l_s$ was used here, $l_s = v_{slip}/\dot{\gamma}$, where $v_{slip}$ is the slip velocity and $\dot{\gamma} = (d\langle v_x \rangle_t/dz)_{bulk}$ is the shear rate within the bulk region of the confined water. The slip velocity was obtained as the difference between $V_{wall}$ and the linear extrapolated bulk velocity at the hydrodynamic wall position (HWP) $V_{HWP}$, i.e., $v_{slip}=V_{wall}-V_{HWP}$ (details about $l_s$ calculation are given in SI section 7). During steady state, for $V_{wall}$ = 50 m/s and 100 m/s, the estimated $l_s$ using APM (30.8 ± 0.7 nm and 34.1 ± 1.0 nm respectively) is similar to that from the conventional method (37.3 ± 8.7 nm and 50.0 ± 13.2 nm respectively) as shown in Fig. 3g, confirming the validity of using APM in nanofluidics.

The dependence of $l_s$ on $v_{slip}$ is a key property in nanofluidics (*43*). Based on transition state theory, $l_s$ is predicted to be a constant at water-carbon surface when $v_{slip}$ is relatively low, e.g., less than 100 m/s (*20*). Such a prediction has been verified by experiments and MD simulations separately. However, the flow rate ranges investigated deviate by large, 20 to 60 μm/s in experiments (*24*) and 3 to 30 m/s in MD simulations (*21*). With the system shown in Fig. 3a, we calculated $l_s$ with $v_{slip}$ from $9.6 \times 10^{-5}$ to 97 m/s using APM during steady state, of which the corresponding flow rate range covers that of nanoconfined fluid in the experiments (*9*) (Fig. 3h). As shown in Fig. 3g, within such a wide range, $l_s$ remains a constant (30.5 ± 1.5 nm). We further compared these values to $l_s$ with $v_{slip}$ = 0. The latter can be estimated using the equilibrium molecular dynamics (EMD) method (*44*) as $l_s = \eta_m/\lambda$, where $\lambda$ is friction coefficient and $\eta_m$ is the viscosity of water at bulk region (0.7 nm away from the graphene). This value can be regarded as the low-speed limit of $l_s$. With EMD, we obtained $\lambda = (1.9 \pm 0.03) \times 10^4$ kg·m$^{-2}$s$^{-1}$, resulting in $l_s = 33 \pm 1$ nm with $\eta_m = (6.3 \pm 0.2) \times 10^{-4}$ kg·m$^{-1}$s$^{-1}$. The agreement between $l_s$ predicted by APM (30.5 ± 1.5 nm) and by EMD (33 ± 1 nm with $v_{slip} = 0$) shows the validity of the APM in low-speed limit. The value of $l_s$ (30.5 ± 1.5 nm) also lies in the range measured experimentally (16 nm for graphene nanochannels (*45*) and 60 nm for graphite nanochannels (*46*)). Together with the comparison to the conventional method, it is reasonable to conclude that the APM method works in a vast range of slip velocity for nanofluidics. As a result, for the dependence of $l_s$ on $v_{slip}$, by bridging the gap spanning over 4 orders between experiments (*24*) and simulations (*21*), we found that $l_s$ is velocity-independent when $v_{slip} < 97$ m/s, which validate the theoretical predictions (*24*).

Besides nanofluidics, the other typical example application of APM is liquid superlubricity. We studied the sliding of water on FDTS (Perfluorodecyltrichlorosilane, $C_{10}H_4Cl_3F_{17}Si$) monolayer (heterogeneous surfaces). Such a system shows good superlubric properties among the many systems of liquid superlubricity (*47*). However, the dependence of friction coefficient $\lambda$ where $\tau = \lambda v_{slip}$ on temperature remains unclear. Such dependence is important not only for the fundamental understanding of the physical mechanisms of slip at the interface, but also for applications like bearings and cooling of electronic devices.

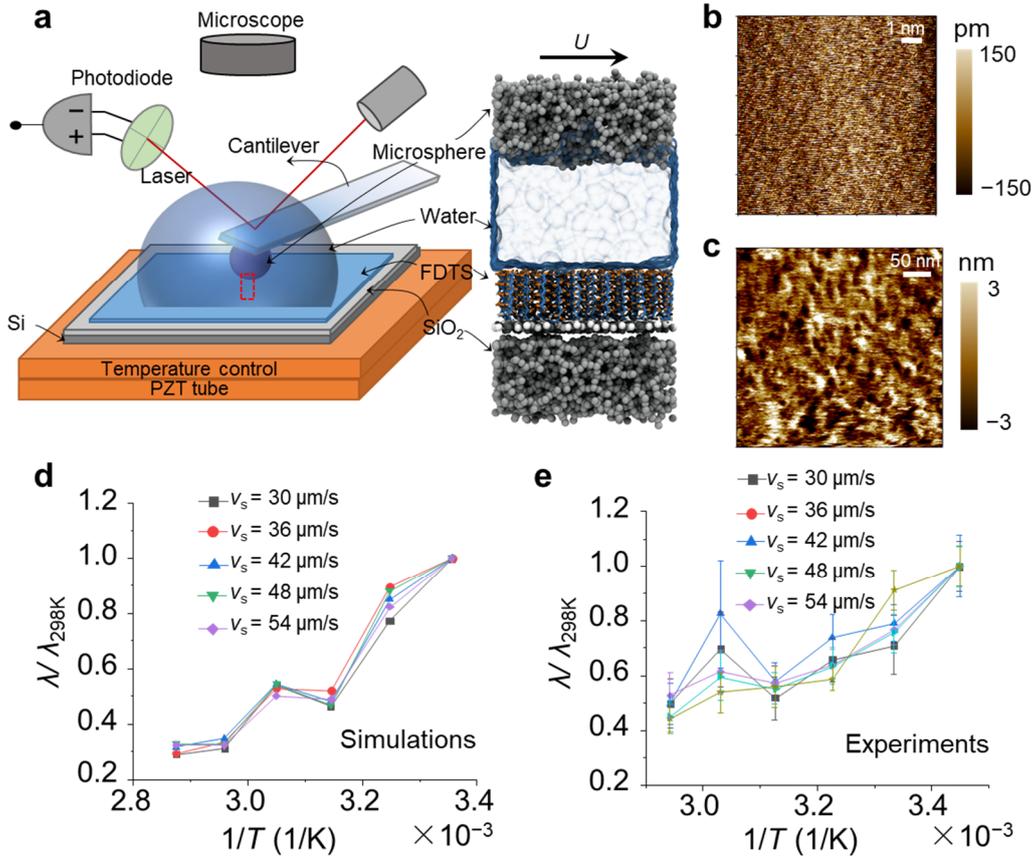

**Figure 4. Direct comparison between the temperature dependence of friction coefficient measured in experiments and predicted by APM. a,** Schematic of the set-up to measure friction coefficient at water-FDTS interface in experiments (left) and MD simulations (right). The area highlighted by the dotted red rectangle is the system simulations considered. **b,** Morphology characterization of the FDTS monolayer surface, with root-mean-square (RMS) roughness of 0.08 nm for a 10×10 nm$^2$ area. **c,** Morphology characterization of the microsphere surface, with RMS roughness of 1.12 nm for a 400×400 nm$^2$ area. **d – e,** Temperature dependence of friction coefficient normalized by the value at 298 K for MD simulations (**d**) and experiments (**e**) under the same shearing velocity ($v_s$) range, i.e., 30 μm/s to 54 μm/s.

Here, we constructed a MD simulation model considered water (a thickness of 3 nm) confined by FDTS monolayer and amorphous silica microsphere, as shown in Fig. 4a. The silica surface, with a lateral size of 4.5×4.5 nm$^2$, sheared the water molecules with a constant velocity $U$. The details of the simulation are presented in SI section 4. By using the APM, the velocity profiles of water flow were obtained with $U$ from 30 μm/s to 54 μm/s and temperature $T$ from 298 K to 348 K (see Fig. S4 in SI section 5). Then the local viscosity distribution was calculated using the same process as that in Fig. 3e (see Fig. S9 in SI section 6). The friction coefficient at water-FDTS interface $\lambda = \eta_m/l_s$ can be calculated with the obtained velocity profiles and viscosity distribution, and its temperature dependence was plotted in Fig. 4d.

To validate the temperature dependence of friction coefficient predicted by MD simulations using APM, a well-designed experiment was performed as shown in Fig. 4a. The self-assembled FDTS monolayer was physically vapor-deposited onto a silica surface. A microsphere was fixed on the atomic force microscope (AFM) cantilever. The morphology of the FDTS monolayer (Fig. 4b) and microsphere (Fig. 4c) indicated clean surfaces. During the measurement, the FDTS monolayer and the microsphere were immersed into degassed deionized water (18.2 MΩ, Hitech Sciencetool). The details of the measuring process are presented in SI section 9. During the measurement, the FDTS surface approached to the microsphere with a constant normal velocity from 50 μm/s to 90 μm/s, resulting a tangential shearing velocity from 30 μm/s to 54 μm/s at the water-FDTS interface (derivation was given in SI section 9). The temperature $T$ from 290 K to 340 K was used in experiment. The same velocity and temperature ranges used in experiments and APM guarantee a direct comparison between their results. The separation $D$ and hydrodynamic resistance $F_D$ between the microsphere and FDTS surface were recorded during the measurement (see Fig. S11 in SI section 9). The interfacial friction coefficient can be calculated from the $F_D$-$D$ curves as suggested in Ref. (*48-50*) with results shown in Fig. 4e.

Across the velocities studied, we observed clear thermolubric phenomena, i.e., the friction coefficient decreases with temperature (Figs. 4d and 4e). With $T$ increases from 298 K to 348 K, by averaging $\lambda$ over the cases for different velocities, the decrease in $\lambda$ is found to be 67% in simulation and 55% by experiment, showing good agreement. This set of results based on a synergetic approach not only shows the thermolubric behavior for water on FDTS unambiguously, but also indicates rich information on atomic scale could be obtained to understand the mechanisms.

## Discussion

With the APM, we show that one can extract an accurate estimate of the dynamical quantity from the huge relative fluctuations for system at nonequilibrium. For the conventional method, its standard error of the estimate is caused by two parts. One is the fluctuation of the initial equilibrium states. The other is the coupling between perturbation and equilibrium states. For APM, by using $\langle\ \rangle_{EC}$, the effect of the fluctuation of the initial equilibrium states on the estimate can be greatly suppressed. Meanwhile, the effect of the coupling between perturbation and equilibrium states on the estimate remains unchanged. Benefit from this property, as shown below, the evolution of the dynamical quantities under external perturbation can be observed clearly.

In Figs. 5a - 5c, with APM we plot the evolution of the velocity fields for water sheared by the graphene sheets using the same model as shown in Fig. 3a. Details of the flow field calculation are given in SI section 5. The streamlines which are a family of curves that are instantaneously tangent to the velocity vector of the flow were computed by using the velocity fields. At any instantaneous moment, the fluctuation of the molecular velocity drives the flow field away from the typical estimation of Stokes flow, of which the streamlines should be parallel with the shearing wall strictly. The steady flow fields (the last column in Fig. 5) which were obtained by sampling the instantaneous flow field every 1 ps, then averaged over 500 ps during the steady state, however, show a good agreement with the prediction of Stokes flow. Compared with the flow field calculated by conventional method (Fig. 5d), it is evident that the APM can show the details of the flow fields evolution at atomic scale, with flow velocity down to a few μm/s.

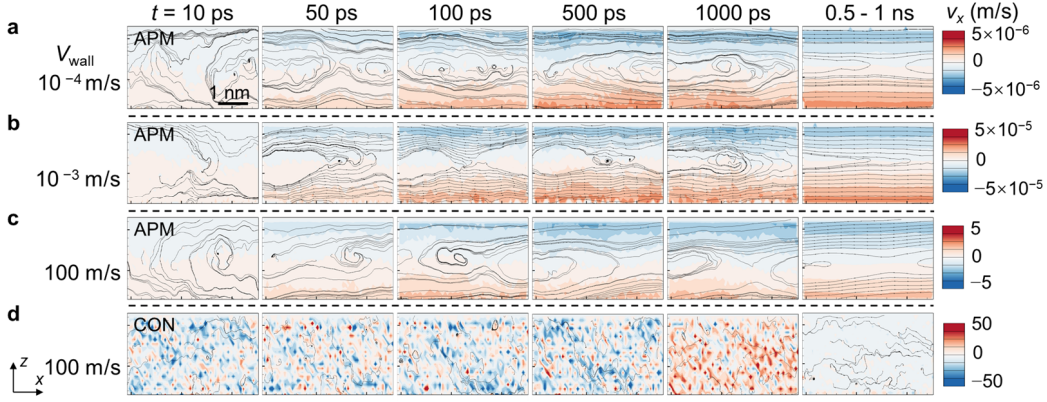

**Figure 5. Evolution of the flow field for water sheared by graphene. a - c,** The flow fields shown in the first three rows were obtained from the APM with $V_{wall} = 10^{-4}$, $10^{-3}$, or 100 m/s. **d,** The flow fields shown in the last row were obtained from the conventional method with $V_{wall} = 100$ m/s. The streamlines which are a family of curves that are instantaneously tangent to the velocity vector of the flow were computed by the velocity field with a grid size of 0.1 nm × 0.1 nm (see SI section 5 for details). Movies S1 - S4 corresponding to (**a - d**) are also provided as SI.

## Conclusion

To conclude, we developed a theory which can give the accurate estimate of dynamical quantities for systems with huge fluctuations at nonequilibrium. This is done by the construction of auxiliary path for each real path in the phase space. The states on auxiliary paths constitute canonical ensemble and share the same macroscopic properties with the initial states of the real path. Taking nanofluidics as the first example, we demonstrated a 12 orders of magnitude increasement with APM in the statistical efficiency for the estimation of flow velocity. As a result, the proposed auxiliary path method can obtain flow fields with a spatial precision of 0.1 nm and temporal resolution of 10 ps, for flow rate down to a few μm/s, which is within the reach of practical experiments (*17, 51*). Indeed, this is validated by the direct comparison between the friction coefficients measured in experiment and estimated by APM in our second example where water is sliding on FDTS surface. Therefore, it is reasonable to expect a series of discoveries for nanofluidics and liquid lubrication with synergetic study by combining APM and experiments together.

On a broader scope, the APM theory only assumes that the system is initially in thermal equilibrium then driven from that equilibrium by an external perturbation, and the dynamics of the system is deterministic. These two general assumptions pose little limitation on the applicable range of the theory. Therefore, the APM could serve as a general approach to provide insights on atomic level under experimental conditions. The specific protocol further assumes that if the system is unperturbed, then it preserves equilibrium ensemble. We expect that with further developments in the construction of the auxiliary path, the gap between atomic simulations and experiments can be filled for fields like sliding between solid surfaces, the mass transport on nanoscale driven by pressure or chemical potential drop, and the multiphase flow under extreme confinements.

**Acknowledgements**

M.M. acknowledges the support from the NSFC (Grant no. 11890673, 11772168 and 51961145304), national supercomputer center in Tianjin and supercomputer Tansuo 100 of Tsinghua University.

# Accurate estimation of dynamical quantities for nonequilibrium nanoscale system

**Supplementary Information**

Zhi Xu[1,2], Han Li[1,2], Ming Ma[1,2]*

[1]Department of Mechanical Engineering, State Key Laboratory of Tribology, Tsinghua University, Beijing 100084, China

[2]Center for Nano and Micro Mechanics, Tsinghua University, Beijing 100084, China

*Email: maming16@tsinghua.edu.cn

In this supplementary information, we provide additional details on certain aspects of the study reported in the manuscript. The following issues are discussed:

1. Relation between relative fluctuation and particle number
2. Derivation of the distribution function for middle states
3. Derivation of the upper limit of the bias
4. Details of MD simulation
5. Post-processing for velocity profile and flow field
6. The calculation process of quantitative viscosity distribution
7. Slip length calculations
8. Computing time estimation
9. The experimental measuring details
10. Description of supporting movies

## 1. Relation between relative fluctuation and particle number

In the main text, we use the relation between relative fluctuation (RF) of the dynamical quantities and the number of particles (*N*) of the system, RF $\propto 1/\sqrt{N}$ (*1*). Here we give the full derivation of this relation and the expression of RF for the centre of mass velocity of a steady water flow.

Imagining that the system we studied is divided into $N_p$ of approximately equal small parts. Then

$$A_i = \frac{1}{N_p} \sum_{j=1}^{N_p} A_{i,j}, \tag{1}$$

where $A_i$ is the intensive quantity related to the *i*th microstate of the system and the quantity $A_{i,j}$ relates to the individual part *j*. Then the fluctuation of *A* can be written as:

$$\sigma_A = \sqrt{\frac{1}{Q}\sum_i^Q (A_i - \bar{A})^2} = \sqrt{\frac{1}{Q}\sum_{i=1}^Q \left(\sum_{j=1}^{N_p}\left(\frac{A_{i,j}}{N_p} - \frac{\bar{A}}{N_p}\right)\right)^2} = \sqrt{\frac{1}{Q}\frac{1}{N_p^2}\sum_{i=1}^Q \left(\sum_{j=1}^{N_p}(A_{i,j} - \bar{A})\right)^2}, \tag{2}$$

where $\bar{A} = \frac{1}{Q}\sum_i^Q A_i$ and *Q* is the number of microstates of the system. Because of the statistical independence of the different parts *j*, the mean value of the products $(A_{i,j} - \bar{A})(A_{i,k} - \bar{A})$, i.e.

$$\frac{1}{Q}\sum_i^Q (A_{i,j} - \bar{A})(A_{i,k} - \bar{A}) = \frac{1}{Q}\sum_i^Q (A_{i,j} - \bar{A}) \times \frac{1}{Q}\sum_i^Q (A_{i,k} - \bar{A}) = 0 \quad (j \neq k), \tag{3}$$

since $\frac{1}{Q}\sum_i^Q (A_{i,j} - \bar{A}) = \frac{1}{Q}\sum_i^Q (A_{i,k} - \bar{A}) = 0$ for intensive quantities *A* with large enough *Q*. Therefore, equation (2) can be further simplified to $\sigma_A = \sqrt{\frac{1}{Q}\frac{1}{N_p^2}\sum_{i=1}^Q \sum_{j=1}^{N_p}(A_{i,j} - \bar{A})^2}$. Substituting equation (1) into the definition of $\bar{A}$, we have $\bar{A} = \frac{1}{Q}\sum_i^Q A_i = \frac{1}{Q}\sum_i^Q \frac{1}{N_p}\sum_{j=1}^{N_p} A_{i,j}$. Hence the relative fluctuation can be expressed as

$$\text{RF} = \frac{\sigma_A}{\bar{A}} = \frac{\sqrt{\frac{1}{Q}\frac{1}{N_p^2}\sum_{i=1}^Q \sum_{j=1}^{N_p}(A_{i,j} - \bar{A})^2}}{\frac{1}{QN_p}\sum_i^Q \sum_{j=1}^{N_p} A_{i,j}} \propto \frac{1}{\sqrt{N_p}}. \tag{4}$$

Considering a homogeneous body to be divided into parts of a given small size, it is clear that the total number of particles *N* in the system will be proportional to the number of parts $N_p$. Hence the result can also be stated that the RF of any intensity quantity $A$ is inversely proportional to the square root of the number of particles in the system, i.e., RF $\propto 1/\sqrt{N}$, as RF $\propto \frac{1}{\sqrt{N_p}}$ and $N \sim N_p$. Note that here we only give the derivation for intensity quantity, for extensive quantity the similar derivation can be done to get the same conclusion.

Next, we give the expression of RF for the centre of mass velocity in $x$ direction $v_x$ for a steady water flow. The canonical ensemble distribution function for system including $N$ water molecules

$$F(\mathbf{r}_1, \ldots, \mathbf{r}_N, \mathbf{p}_1, \ldots, \mathbf{p}_N) = \frac{e^{-\beta H(\mathbf{r}_1, \ldots, \mathbf{r}_N, \mathbf{p}_1, \ldots, \mathbf{p}_N)}}{\int d\mathbf{r}^N d\mathbf{p}^N e^{-\beta H(\mathbf{r}_1, \ldots, \mathbf{r}_N, \mathbf{p}_1, \ldots, \mathbf{p}_N)}}, \tag{5}$$

where $\mathbf{r}$ and $\mathbf{p}$ are the coordinate and momentum respectively, $\beta = 1/k_B T$, $H(\mathbf{r}_1, \ldots, \mathbf{r}_N, \mathbf{p}_1, \ldots, \mathbf{p}_N) = \sum_{i=1}^{N} \sum_{\alpha=1}^{3} \frac{p_{i\alpha}^2}{2m_{\text{water}}} + U(\mathbf{r}_1, \ldots, \mathbf{r}_N)$, $p_{i\alpha}$ is the momentum of particle $i$ in $\alpha$ direction, $m_{\text{water}}$ is the mass of individual water molecule and $U(\mathbf{r}_1, \ldots, \mathbf{r}_N)$ is the total potential energy of the system. To obtain the distribution function of the momentum vector $\mathbf{P}_i$ in $x$ direction, $p_{ix}$, the other variables in $F(\mathbf{r}_1, \ldots, \mathbf{r}_N, \mathbf{p}_1, \ldots, \mathbf{p}_N)$ are integrated. Then $F(p_{ix}) = \frac{e^{-\frac{\beta}{2m_{\text{water}}}(p_{ix}^2)}}{\int_{-\infty}^{+\infty} dp_x e^{-\frac{\beta}{2m_{\text{water}}}(p_{ix}^2)}}$ and $\int_{-\infty}^{+\infty} dp_x \, e^{-\frac{\beta}{2m_{\text{water}}}(p_{ix}^2)} = \left(\frac{2\pi m_{\text{water}}}{\beta}\right)^{1/2}$. Therefore, the distribution function $F(p_{ix})$ can be expressed as

$$F(p_{ix}) = \left(\frac{\beta}{2\pi m_{\text{water}}}\right)^{\frac{1}{2}} e^{-\frac{\beta}{2m_{\text{water}}}(p_{ix}^2)}. \tag{6}$$

As a result, $p_{ix}$ satisfies a normal distribution with a mean of zero and a standard deviation of $\sqrt{m_{\text{water}}/\beta}$. Therefore, the centre of mass velocity along the $x$ direction $v_{\text{com}} = \frac{1}{N} \sum_{i=1}^{N} \frac{p_{ix}}{m}$ has a relative fluctuation RF $= \sigma_{v_{\text{com}}}/v_{\text{com}}$ where $\sigma_{v_{\text{com}}} = \sqrt{1/\beta N m_{\text{water}}}$.

## 2. Derivation of the distribution function for middle states

In the main text, we present a specific protocol to generate states on the auxiliary path. To prove that the states meet the requirements of auxiliary path, we use the distribution function of middle states. The full derivation is presented here.

In Fig. 2a of the main text, the schematic of the specific protocol for auxiliary path method (APM) is presented. For states $x'_{\delta t}$, as they are propagated from $x''_0$ without any perturbation, they constitute a canonical ensemble. In other words, the probability of states $x'_{\delta t}$, $f(x'_{\delta t}) = \frac{e^{-\beta H(x'_{\delta t})}}{\sum_i e^{-\beta H(x'_{\delta t})}}$ where $i$ denotes the $i$th path in the path ensemble. Further, the essence of this protocol is a one-to-one correspondence between the real path and the auxiliary path, hence each middle state $x''_{\delta t}$ is determined by the auxiliary state $x'_{\delta t}$ and its unique corresponding real state, which means $f(x''_{\delta t}) = f(x'_{\delta t})$, i.e.,

$$f(x''_{\delta t}) = \frac{e^{-\beta H(x'_{\delta t})}}{\sum_i e^{-\beta H(x'_{\delta t})}}$$

$$= \frac{e^{-\beta H(x''_{\delta t})} e^{-\beta\left(H(x'_{\delta t})-H(x''_{\delta t})\right)}}{\sum_i e^{-\beta H(x''_{\delta t})} e^{-\beta\left(H(x'_{\delta t})-H(x''_{\delta t})\right)}} = \frac{e^{-\beta H(x''_{\delta t})} e^{-\beta \Delta H_i(\delta t)}}{\sum_i e^{-\beta H(x''_{\delta t})} e^{-\beta \Delta H_i(\delta t)}}. \tag{7}$$

In order to obtain the relation between $f(x''_{\delta t})$ and the canonical ensemble distribution function $e^{-\beta H(x''_{\delta t})}/\sum_i e^{-\beta H(x''_{\delta t})}$ of the middle states, we first determine the range of $f(x''_{\delta t})$.

$$\frac{\min_i\{e^{-\beta \Delta H_i(\delta t)}\} e^{-\beta H(x''_{\delta t})}}{\max_i\{e^{-\beta \Delta H_i(\delta t)}\} \sum_i e^{-\beta H(x''_{\delta t})}} \leq f(x''_{\delta t}) \leq \frac{\max_i\{e^{-\beta \Delta H_i(\delta t)}\} e^{-\beta H(x''_{\delta t})}}{\min_i\{e^{-\beta \Delta H_i(\delta t)}\} \sum_i e^{-\beta H(x''_{\delta t})}}. \tag{8}$$

Then the formula (8) is divided by $e^{-\beta H(x''_{\delta t})}/\sum_i e^{-\beta H(x''_{\delta t})}$,

$$\frac{1}{a} \leq \frac{f(x''_{\delta t})}{\frac{e^{-\beta H(x''_{\delta t})}}{\sum_i e^{-\beta H(x''_{\delta t})}}} \leq a, \quad a = \frac{\max_i\{e^{-\beta \Delta H_i(\delta t)}\}}{\min_i\{e^{-\beta \Delta H_i(\delta t)}\}}. \tag{9}$$

We simplify $a$ as $e^{\lambda(\delta t)}$ where $\lambda(\delta t) = \beta\left(\max_i\{\Delta H_i(\delta t)\} - \min_i\{\Delta H_i(\delta t)\}\right)$. When $\lambda(\delta t) \to 0$,

$$\lim_{\lambda(\delta t) \to 0} a = \lim_{\lambda(\delta t) \to 0} e^{\lambda(\delta t)} = 1. \tag{10}$$

Combined with formula (9), we can derive that

$$f(x''_{\delta t}) = \frac{e^{-\beta H(x''_{\delta t})}}{\sum_i e^{-\beta H(x''_{\delta t})}}. \tag{11}$$

Overall, we can conclude that the states $x''_{\delta t}$ satisfy canonical ensemble distribution function, thus constitute a canonical ensemble when $\lambda(\delta t) \to 0$. The function $\lambda(t)$ is a dimensionless quantity that varies in time $t$, when $\lambda(t) \to 0$ is valid for every moment $t$, the construct states $x''_t$ will form an auxiliary path.

## 3. Derivation of the upper limit of the bias

In this section, we give the analysis of the bias for the auxiliary states away from the canonical ensemble due to the finite small $\lambda$ in the practical construction process.

The auxiliary path requires that the auxiliary states $x'_t$ constitute a canonical ensemble at any time $t$, thus $\lim_{Q \to \infty} \frac{1}{Q} \sum_i^Q A'_i(t)$ will be equal to the canonical ensemble average $\langle A \rangle_{\mathrm{EC}}$, where $Q$ is the number of paths in the path ensemble. In the main text, we give the construction process of auxiliary states which constitute canonical ensembles under the condition $\lambda(t) \to 0, \lambda(t) = \beta\left(\max_i\{\Delta H_i(t)\} - \min_i\{\Delta H_i(t)\}\right)$, $i$ denotes the $i$th path in the path ensemble. When $\lambda = 0$, $\lim_{Q \to \infty} \frac{1}{Q} \sum_i^Q A'_i(t) = \langle A \rangle_{\mathrm{EC}}$. However, the finite small $\lambda$ in the practical construction process will cause a bias of the auxiliary states away from the canonical ensemble. The bias can be expressed as $\phi(t) = \lim_{Q \to \infty} \frac{1}{Q} \sum_i^Q (A'_i(t) - \langle A \rangle_{\mathrm{EC}})$. $\phi(t)$ is a sum of two parts, one part is due to the construction from auxiliary states $x'_t$ to middle states $x''_t$, the other is due to the relaxation from middle states $x''_t$ to auxiliary states $x'_{t+\delta t}$ at the next timestep. The first part $\eta(t)$ is the bias of $\lim_{Q \to \infty} \frac{1}{Q} \sum_i^Q A''_i(t)$ from $\lim_{Q \to \infty} \frac{1}{Q} \sum_i^Q A'_i(t)$, i.e., $\eta(t) = \lim_{Q \to \infty} \frac{1}{Q} \sum_i^Q (A''_i(t) - A'_i(t))$. The auxiliary states $x'_{t+\delta t}$ are obtained after a relaxation of time $\delta t$ from $x''_t$, hence $\phi(t + \delta t) = \alpha(\phi(t) + \eta(t))$, where $\alpha = e^{-\delta t/\tau}$ and $\tau$ is the relaxation time of the quantity $A$ for a given system based on Onsager's regression hypothesis (2). Considering from $t = 0$ to $t = C\delta t$ where $C$ is the number of timesteps accumulated, the bias $\phi(C\delta t)$ can be derived as

$$\phi(C\delta t) = \alpha\left(\ldots\left(\alpha\big(\alpha\eta(\delta t) + \eta(2\delta t)\big) + \eta(3\delta t)\right)\ldots + \eta\big((C-1)\delta t\big)\right)$$

$$= \sum_{j=1}^{C-1} \alpha^{C-j} \eta(j\delta t). \tag{12}$$

Next, we give an estimate of the upper limit of $\eta(t)$. Notice that $e^{-\lambda(\delta t)} \leq f(x''_{\delta t}) / \dfrac{e^{-\beta H(x''_{\delta t})}}{\sum_i e^{-\beta H(x''_{\delta t})}} \leq e^{\lambda(\delta t)}$ is valid as shown in the Supplementary Information (SI) Section 3, where $\lambda(\delta t) = \beta \left( \max_i \{\Delta H_i(\delta t)\} - \min_i \{\Delta H_i(\delta t)\} \right)$. In the first order approximation of $e^{-\lambda(\delta t)}$ and $e^{\lambda(\delta t)}$ with $\lambda(\delta t) \to 0$, $(1-\lambda)\dfrac{e^{-\beta H(x''_{\delta t})}}{\sum_i e^{-\beta H(x''_{\delta t})}} \leq f(x''_{\delta t}) \leq (1+\lambda)\dfrac{e^{-\beta H(x''_{\delta t})}}{\sum_i e^{-\beta H(x''_{\delta t})}}$. It means that the probability of middle states $f(x''_{\delta t})$ is in the range of $(1-\lambda(\delta t), 1+\lambda(\delta t))$ times the probability of canonical ensemble. The upper limit of $\eta(\delta t)$ occurs when the probability of states that make the corresponding $A''(\delta t)$ larger than $\langle A \rangle_{\mathrm{EC}}$ change with the upper bound by $1 + \lambda(\delta t)$ times, and the probability of states that make $A''(\delta t)$ smaller than $\langle A \rangle_{\mathrm{EC}}$ change with the lower bound by $1 - \lambda(\delta t)$ times. As a result, the max value of $\eta(\delta t)$ can be estimated as $\lambda(\delta t)\sigma_A$. Similar analysis can be done for the subsequent time steps, thus we can estimate the upper limit of $\eta(t)$ with $\lambda(t)\sigma_A$, resulting in

$$\phi(C\delta t) = \sum_{j=1}^{C-1} \alpha^{C-j} \eta(j\delta t) < \sigma_A \sum_{j=1}^{C-1} \alpha^{C-j} \lambda(j\delta t). \tag{13}$$

Therefore, the upper limit of the bias for the auxiliary states away from the canonical ensemble is $\sigma_A \sum_{j=1}^{C-1} \alpha^{C-j} \lambda(j\delta t)$.

The randomness of the bias for the auxiliary states away from the canonical ensemble is analyzed here. In fact, $\eta(t)$ can be positive or negative with no distinct preference. Thus $\phi(C\delta t)$, which is the sum of $\alpha^{C-j}\eta(j\delta t)$, can be considered reasonably random with the max absolute value $\sigma_A \sum_{j=1}^{C-1} \alpha^{C-j} \lambda(j\delta t)$. Further,

$$\sigma_A \sum_{j=1}^{C-1} \alpha^{C-j} \lambda(j\delta t) < \sigma_A \lambda_{\max} \sum_{j=1}^{C-1} \alpha^{C-j} = \sigma_A \lambda_{\max} \frac{\alpha(1-\alpha^{C-1})}{1-\alpha}, \tag{14}$$

where $\lambda_{\max}$ is the max value of $\lambda(t)$ from $t=0$ to $t=C\delta t$. It means that the upper limit $\sigma_A \sum_{j=1}^{C-1} \alpha^{C-j} \lambda(j\delta t)$ will be smaller than $\sigma_A \lambda_{max} \frac{\alpha}{1-\alpha}$ as time accumulates.

## 4. Details of MD simulations

In this section, we give the details of the molecular dynamics (MD) simulations for both two systems mentioned in the main text, system for water sheared by graphene (a) and system for water on FDTS monolayer sheared by silica (b).

**a.** Simulation system for water sheared by graphene

The SPC/Fw force field (*3*) was used for water molecules, which reproduces the dynamical properties of water close to the experimental values. The graphene sheets were modeled using the second-generation REBO potential (*4*), which is widely used for graphene. The interaction between water molecules and carbon atoms of graphene was modeled using the Lennard-Jones potential with parameters from Ref. (*5*), with $\varepsilon_{CO} = 4.063$ meV and $\sigma_{CO} = 0.319$ nm. It corresponds to a contact angle $\theta$ of 86° for water on graphene. Electrostatic interaction between water molecules was modeled using the modified Wolf potential with damping parameter $\alpha = 0.2$ (*6*). With a long enough cutoff and small enough $\alpha$, the energy and force calculated by the Wolf summation method approach those of the Ewald sum. The interlayer potential (ILP) (*7*) was applied for carbon atoms belonging to the different graphene layers. All Lennard-Jones interactions and electrostatic interactions were truncated at a distance of 1.2 nm and then linearly smooth to zero (*6, 8*), for a continuous energy and force at the cutoff radius. All the MD simulations were carried out using a time step of 1 fs.

The simulation procedure included the following steps:

1) Obtaining the correct fluid density
2) Equilibrating the system
3) Setting the flow rate to be zero
4) Starting shearing and collecting data

**a.1** Obtaining the correct fluid density

At the beginning of each simulation, the bottom two graphene sheets (1st and 2nd layers) were fixed, the top two graphene sheets (3rd and 4th layers) were constrained to move only in the perpendicular direction as a rigid body under a fixed pressure of 1 atm as shown in Fig. S1a. The fluid was maintained at a room temperature of 298 K using Nosé-Hoover chain thermostat. The simulation lasted for 50 ps, which is long enough to get a stable distance between innermost graphene sheets and thus a correct fluid density. This procedure has been shown to produce a reasonable density of fluid confined in graphene sheets (9).

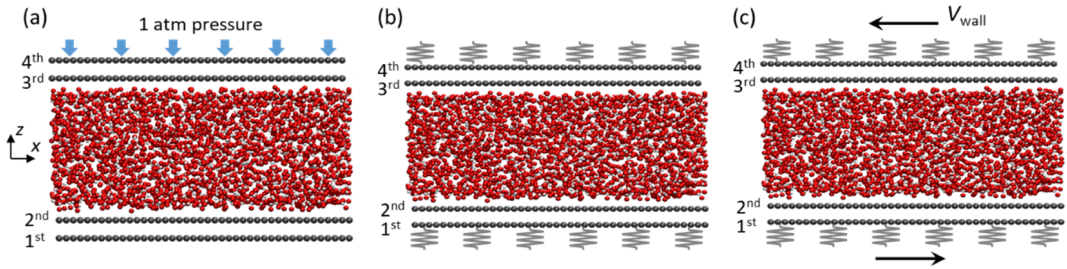

**Figure S1. Schematic of simulation procedures**. **a,** Obtaining the correct fluid density. **b,** Equilibrating the system. **c,** Starting shearing the fluid.

**a.2** Equilibrating the system

The normal force applied on the top graphene layer which was used to generate a fixed pressure of 1 atm was removed, and the constraints of the bottom two graphene sheets were also removed. Meanwhile, a spring of 2.7 N/m was added to the carbon atoms in the 1st and 4th layer of graphene (Fig. S1b), i.e., the outermost carbon atoms. Temperature of 298 K was maintained by applying the Nosé-Hoover chain thermostat to the outermost carbon atoms (10) only, so that the heat produced by the slip and viscous friction in the fluid was conducted away through the graphene as is done in a real experiment. The equilibration lasted for 50 ps.

**a.3** Setting the flow rate to be zero

The centre of mass velocity of the fluid was set to be zero before shearing. This is to make sure that no net flow exists when there is no shearing.

**a.4** Starting shearing and collecting data

The wall started moving in opposite directions along $x$-axis with a constant velocity $V_{wall}$ as shown in Fig. S1c. Both the real states and the auxiliary states were recorded during the simulations.

**b.** Simulation system for water on FDTS monolayer sheared by silica

As shown in Fig. S2, a model containing a self-assembled FDTS monolayer covered on amorphous $SiO_2$ substrate was made as suggested in Ref. (*11*). The amorphous $SiO_2$ substrate was created from annealing (*12*), with a lateral size of 4.5 × 4.5 nm$^2$. FDTS monolayer covered the substrate with a coverage of 5 molecules/nm$^2$, and were perpendicular to the substate. The microsphere using in experiment is made of glass and mainly amorphous silica, thus was represented by an amorphous $SiO_2$ surface here in the simulations. The system was then packed with water molecules in the middle of the FDTS monolayer and $SiO_2$ surface. AMBER force field was used for FDTS molecules (*11*). SPC/E force field (*13*) was used for water molecules. CLAYFF force field (*14*) was used for both $SiO_2$ substrate and $SiO_2$ surface. The interaction between these molecules was described using the Lennard–Jones force field. Lorentz–Berthelot mixing rules were used for the inter-molecular interactions. Long-range Columbic interactions were computed using the particle–particle particle–mesh (PPPM) algorithm. All Lennard-Jones interactions were truncated at a distance of 1.2 nm and then linearly smooth to zero (*15*), for a continuous energy and force at the cutoff radius. Periodic boundary conditions in all directions were implemented. All the MD simulations were carried out using a time step of 1 fs.

The simulation procedure included the following steps:
1) Obtaining the correct water density
2) Equilibrating the system
3) Setting the flow rate to be zero
4) Starting shearing and collecting data

**b.1** Obtaining the correct water density

At the beginning of each simulation, the $SiO_2$ substrate was fixed, the top $SiO_2$ surface was constrained to move only in the perpendicular direction as a rigid body under a fixed normal pressure of 1 atm, as shown in Fig. S2a. The water and FDTS molecules were maintained at a temperature from 298 K to 348 K using Langevin thermostat. The simulation lasted for 50 ps, which is long enough to get a stable distance between FDTS and SiO$_2$ surfaces and thus a correct water density.

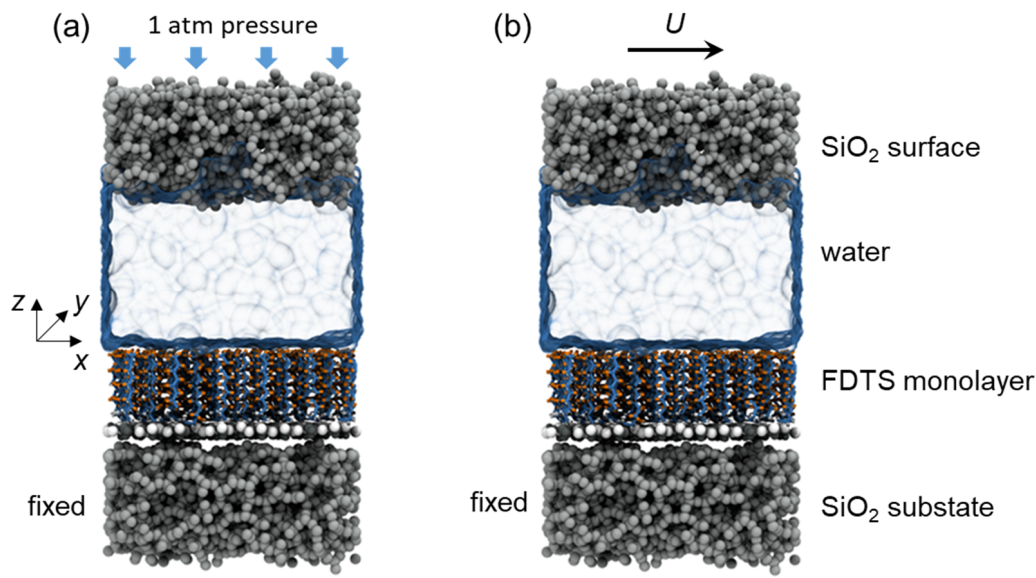

**Figure S2. Schematic of simulation procedures**. **a,** Obtaining the correct water density. **b,** Starting shearing the water.

**b.2** Equilibrating the system

The normal force applied on the top SiO$_2$ which was used to generate a fixed pressure of 1 atm was removed. The SiO$_2$ surface and SiO$_2$ substrate were fixed. Langevin thermostat was set to the FDTS monolayer (*10*) only, so that the heat produced by the slip and viscous friction in the fluid was conducted away through the FDTS monolayer as is done in a real experiment. The equilibration lasted for 50 ps.

**b.3** Setting the flow rate to be zero

The centre of mass velocity of the water was set to be zero before shearing. This is to make sure that no net flow exists when there is no shearing.

**b.4** Starting shearing and collecting data

As shown in Fig. S2b, the SiO$_2$ surface started moving in the directions along *x*-axis with a constant velocity *U*. Both the real states and the auxiliary states were recorded during the simulations.

## 5. Post-processing for velocity profile and flow field

In this section, we first give the detailed data processing for the velocity profile of water flow with data from both APM and the conventional method (CON). Then we give the procedure to plot the flow fields for water sheared by graphene with two methods (APM and CON).

**a. Data processing for velocity profile**

Ten independent simulations ($m$ = 10) were carried for each given shearing velocity to obtain the velocity profile. Each simulation lasted for 1000 ps with a time step of 1 fs. Sampling over $\mathbf{v}_i(t)$ (real states) and $\mathbf{v}'_i(t)$ (auxiliary states) was conducted every 1 ps. For APM, the velocity field of water $\langle \mathbf{v} \rangle_t = \lim_{Q \to \infty} \frac{1}{Q} \sum_{i=1}^{Q} (\mathbf{v}_i(t) - \mathbf{v}'_i(t)) + \langle \mathbf{v} \rangle_{EC}$ where $\mathbf{v}_i(t)$ is the velocity field of state $x_t$ on the $i$th real path and $\mathbf{v}'_i(t)$ is the velocity field of state $x'_t$ on the corresponding auxiliary path. The equilibrium ensemble average of velocity field $\langle \mathbf{v} \rangle_{EC} = 0$. For CON, $\langle \mathbf{v} \rangle_t = \lim_{Q \to \infty} \frac{1}{Q} \sum_{i=1}^{Q} \mathbf{v}_i(t)$.

For the steady velocity profile of water flow, data in the last 500 ps of each simulation was used ($C_s$ = 500). These amount to a total number of samples of 5000 ($m$ = 10, $C_s$ = 500). To resolve the profile, the bin size in the $z$ direction was defined as 0.5 Å. For APM, we averaged $v_x - v'_x$ and the coordinate for atoms in the same bin over 5000 samples. Then the velocity profiles of water flow for system that water sheared by graphene were obtained as shown in Figs. S3a – S3f, corresponding to $V_{wall}$ from $10^{-3}$ m/s to 100 m/s. The velocity profiles of water flow for system that water on FDTS monolayer sheared by $SiO_2$ were obtained as shown in Figs. S4, corresponding to $U$ from 30 μm/s to 54 μm/s.

For CON, the same data processing with $v_x$ was used to obtain the velocity profiles of water flow for system that water sheared by graphene, the velocity profiles were shown in Figs. S5a – S5d corresponding to $V_{wall}$ from 20 m/s to 200 m/s.

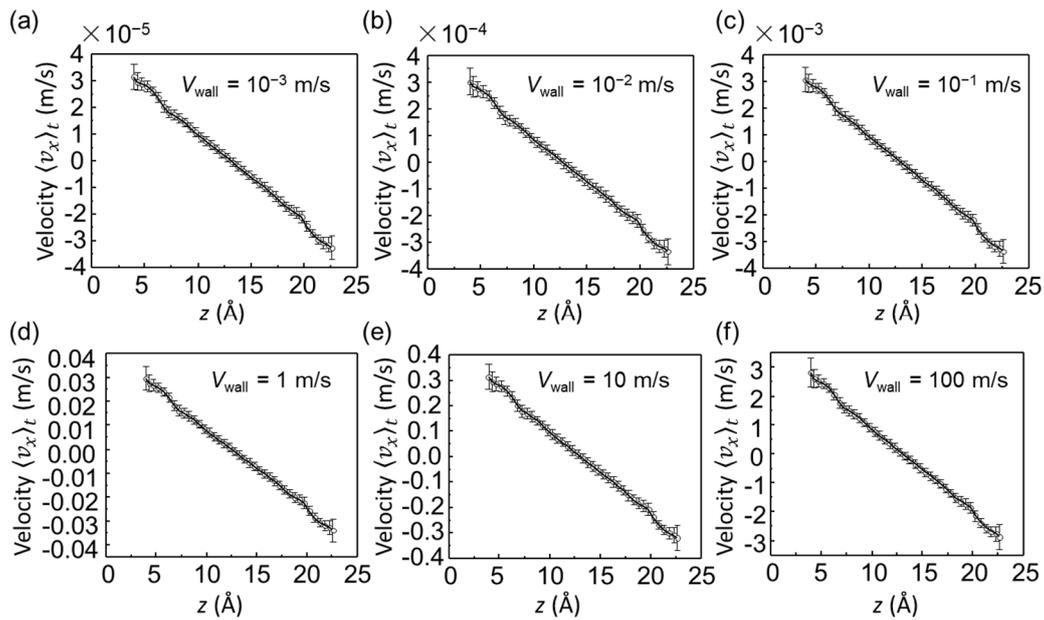

**Figure S3. a-f,** Velocity profiles of water flow obtained from the auxiliary path method with $V_{wall}$ = $10^{-3}$, $10^{-2}$, $10^{-1}$, 1, 10, or 100 m/s.

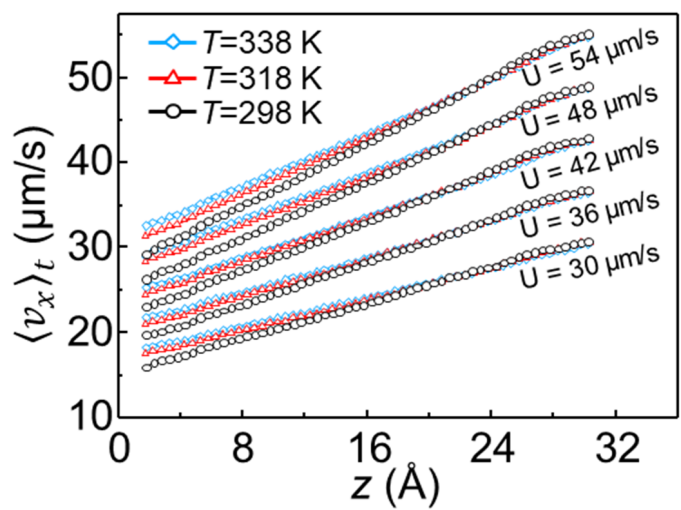

**Figure S4,** Velocity profiles of water flow obtained from the auxiliary path method with $U$ from 30 μm/s to 54 μm/s.

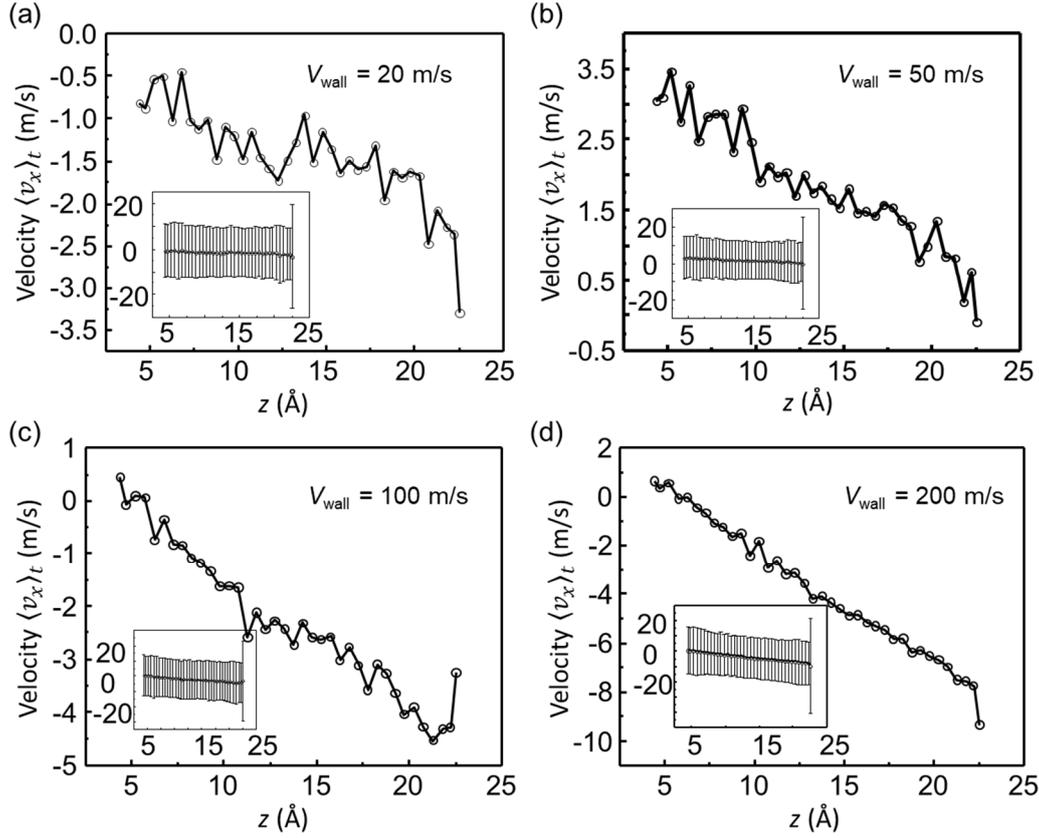

**Figure S5. a-d,** Velocity profiles of water flow obtained from the conventional method with $V_{wall}$ = from 20, 50, 100, or 200 m/s. The insets include the standard error of the velocity.

**b. Data processing for flow field of water sheared by graphene**

For each given $V_{wall}$, one simulation was used to obtain the flow field for both APM and CON. Each simulation lasted for 1000 ps with a time step of 1 fs. Sampling over $\mathbf{v}_i(t)$ (real states) and $\mathbf{v}'_i(t)$ (auxiliary states) was conducted every 1 ps.

For APM, to obtain the instantaneous flow field, the grid size in the x-z plane was defined as 0.1 nm × 0.1 nm. $v_x - v'_x$ and the coordinate of atoms in the same grid were averaged over 10 ps. The flow fields for water sheared by graphene with $V_{wall}$ from $10^{-4}$ m/s to 100 m/s were shown in Figs. S6a – S6c.

For CON, to obtain the instantaneous flow field, the grid size in the x-z plane was defined as 0.3 nm × 0.3 nm. $v_x$ and the coordinate of atoms in the same grid were averaged over 10 ps. The flow field for water sheared by graphene with $V_{wall}$ = 100 m/s was shown in Fig. S6d.

For the streamlines in Fig. 5 of the main text which are a family of curves that are instantaneously tangent to the velocity vector of the flow, they were plotted using an adaptive step-size, trapezoidal integration algorithm. This creates the streamlines by moving it in a series of small steps from the starting point in the direction of the local velocity vector. The step size was automatically adjusted based on the local cell shape and vector field variation.

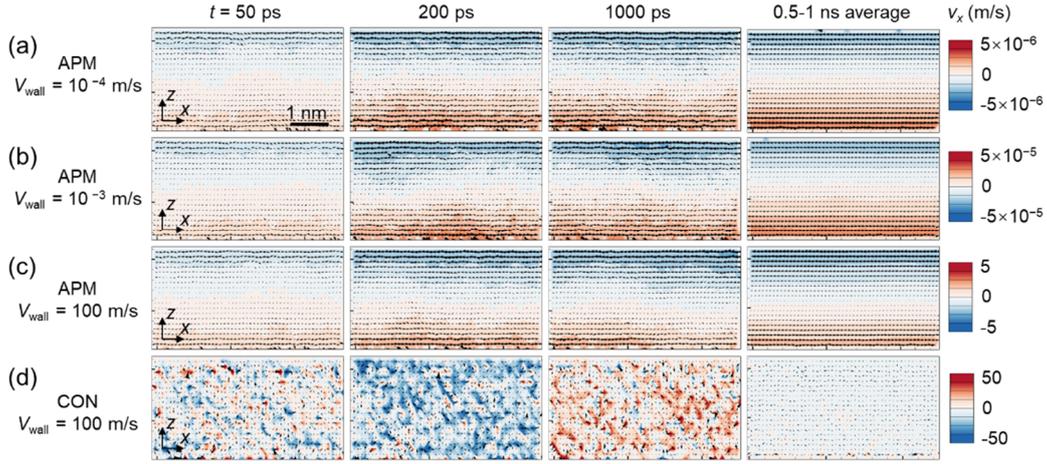

**Figure S6. Evolution of the flow field for water sheared by graphene. a - c,** The flow fields shown in the first three rows were obtained from the APM with $V_{wall} = 10^{-4}$, $10^{-3}$, or 100 m/s. **d,** The flow fields shown in the last row were obtained from the conventional method with $V_{wall} = 100$ m/s. The length of the arrows in the same row is proportional to the magnitude of the velocity, and the direction of the arrows corresponds to the direction of the velocity.

## 6. The calculation process of quantitative viscosity distribution

In this section, we give the calculation process used in Fig. 3e of the main text to achieve the quantitative local viscosity distribution for nanoconfined water. The calculation process is divided into three parts, (a) obtaining relative local viscosity distribution, (b) deriving the relation between local viscosity and viscosity, and (c) calculating the quantitative local viscosity distribution.

**a.** Obtaining relative local viscosity distribution

The local viscosity $\eta(z) = \tau / \frac{d\langle v_x \rangle_t}{dz}(z)$, where $\tau$ is the shear stress that holds constant in the stable flows. From this definition of local viscosity, we can conclude that $\eta$ is inverse proportional to $\frac{d\langle v_x\rangle_t}{dz}(z)$. $\tau = \eta_m \frac{d\langle v_x\rangle_t}{dz}$ ($z = 1.3$ nm) where $\eta_m$ is the local viscosity of the middle region ($z_m = 1.3$ nm) water. Thus, the local viscosity $\eta(z)$ can be expressed as

$$\eta(z) = \eta_m \frac{\frac{d\langle v_x\rangle_t}{dz}|_{z=z_m}}{\frac{d\langle v_x\rangle_t}{dz}|_z}. \tag{15}$$

Given the velocity profile in Fig. S7a, $\eta(z)/\eta_m$ was plotted in Fig. S7b.

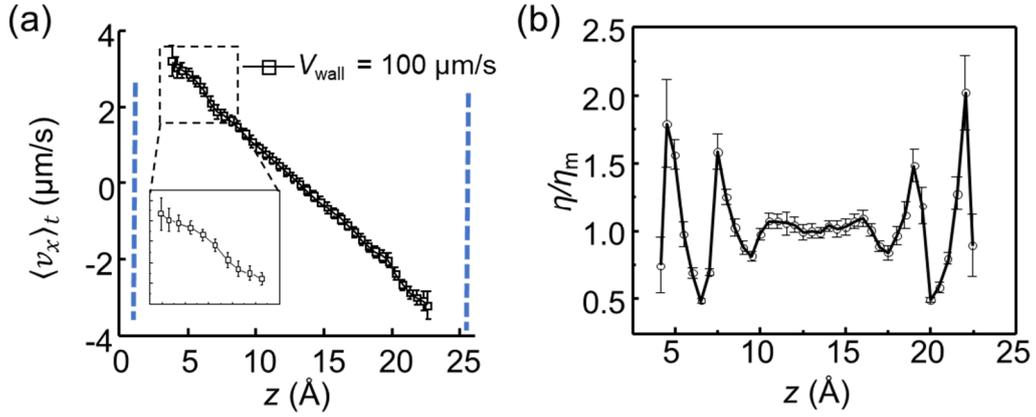

**Figure S7. The relative local viscosity distribution for water flow of the system water sheared by graphene. a,** The velocity profile for water flow. **b,** The relative local viscosity distribution for water.

**b. Deriving the relation between local viscosity and viscosity**

According to Green-Kubo relations, the microscopic definition of viscosity can be expressed as:

$$\eta_{\text{eff}} = \frac{V}{K_B T} \int_0^\infty dt\, \langle P_{xy}(0) P_{xy}(t)\rangle_{\text{EC}}, \tag{16}$$

where $V$ is the volume, $K_B$ is the Boltzmann constant, $T$ is the temperature, $P_{xy}$ is the pressure tensor component, $t$ is time and $\langle\,\rangle_{\text{EC}}$ is the equilibrium ensemble average. For a system consist of $N$ particles,

$$P_{xy}(\mathbf{r},\mathbf{p}) = \frac{1}{V}\sum_{i=1}^{N}\left[\frac{(\mathbf{p}_i\cdot\hat{\mathbf{e}}_x)(\mathbf{p}_i\cdot\hat{\mathbf{e}}_y)}{m_i} + (\mathbf{r}_i\cdot\hat{\mathbf{e}}_x)(\mathbf{F}_i\cdot\hat{\mathbf{e}}_y)\right] = \frac{1}{V}\sum_{i=1}^{N}\sigma_{xy,i}, \tag{17}$$

where $\sigma_{xy,i}$ is the stress tensor component of particle $i$. We divided the system into $N_b$ boxes of the same volume $V_b$,

$$P_{xy} = \frac{1}{V}\sum_{k=1}^{N_b} \sigma_{xy,k} = \frac{V_b}{V}\sum_{k=1}^{N_b} P_{xy,k}, \qquad (18)$$

$\sigma_{xy,k} = \sum_{i=1}^{N_k} \sigma_{xy,i}$, $N_k$ is the number of particles in the $k^{\text{th}}$ box, and $P_{xy,k} = \frac{\sigma_{xy,k}}{V_b}$.

Substituting equation (18) into equation (16) gives

$$\eta_{\text{eff}} = \frac{V}{K_B T}\int_0^\infty dt \left\langle \left(\frac{V_b}{V}\sum_{i=1}^{N_b} P_{xy,i}(0)\right)\left(\frac{V_b}{V}\sum_{j=1}^{N_b} P_{xy,j}(t)\right)\right\rangle_{\text{EC}}$$

$$= \frac{V_b^2}{V K_B T}\int_0^\infty dt \left\langle \left(\sum_{i=1}^{N_b} P_{xy,i}(0)\right)\left(\sum_{j=1}^{N_b} P_{xy,j}(t)\right)\right\rangle_{\text{EC}}. \qquad (19)$$

At equilibrium, we assume that the press-tensor component of the different boxes is independent, which means

$$\langle P_{xy,i}(0) P_{xy,j}(t)\rangle_{\text{EC}} = 0, \text{ for } i \neq j. \qquad (20)$$

Substituting equation (20) into equation (19) gives

$$\eta_{\text{eff}} = \frac{V_b^2}{V K_B T}\sum_{i=1}^{N_b}\int_0^\infty dt \langle P_{xy,i}(0) P_{xy,i}(t)\rangle_{\text{EC}}$$

$$= \frac{V_b}{V}\sum_{i=1}^{N_b} \frac{V_b}{K_B T}\int_0^\infty dt \langle P_{xy,i}(0) P_{xy,i}(t)\rangle_{\text{EC}} = \frac{1}{N_b}\sum_{i=1}^{N_b} \eta_i, \qquad (21)$$

where $\eta_i$ is the local viscosity of box $i$.

c. Calculating the quantitative local viscosity distribution

Combining equation (15) and equation (21),

$$\eta_{\text{eff}} = \frac{1}{N_b}\sum_{z=1}^{N_b} \eta(z) = \frac{1}{N_b}\sum_{z=1}^{N_b} \eta_m \frac{\frac{d\langle v_x\rangle_t}{dz}|_{z=z_m}}{\frac{d\langle v_x\rangle_t}{dz}|_z}. \qquad (22)$$

$\eta_m$ can be calculated as

$$\eta_m = \frac{\eta_{\text{eff}} N_b}{\sum_{z=1}^{N_b} \frac{\frac{d\langle v_x\rangle_t}{dz}|_{z=z_m}}{\frac{d\langle v_x\rangle_t}{dz}|_z}}. \qquad (23)$$

In summary, a standard process should be carried out to achieve the quantitative calculation of local viscosity distribution:

1. With a known velocity profile, calculating relative local viscosity distribution using equation (15). Only the parameter $\eta_m$ remains unknown.
2. Calculating the viscosity of the whole system using equation (16).
3. Obtaining $\eta_m$ using equation (23).

For water sheared by graphene, we first calculated the relative viscosity distribution as shown in Fig. S7b. Then the viscosity for the whole water molecules was calculated as shown in Fig. S8a, as a results, $\eta_{\text{eff}} = 0.64$ mPa·s. Further, we obtained $\eta_m = 0.62$ mPa·s using equation (23), the quantitative local viscosity distribution was plotted in Fig. S8b. The viscosity of bulk water was calculated using Green-Kubo relation with the same data processing, resulting in $\eta_{\text{bulk}} = 0.53$ mPa·s. The match of $\eta_m$ and $\eta_{bulk}$ means that the water at the middel region has the similar properties as that of bulk water.

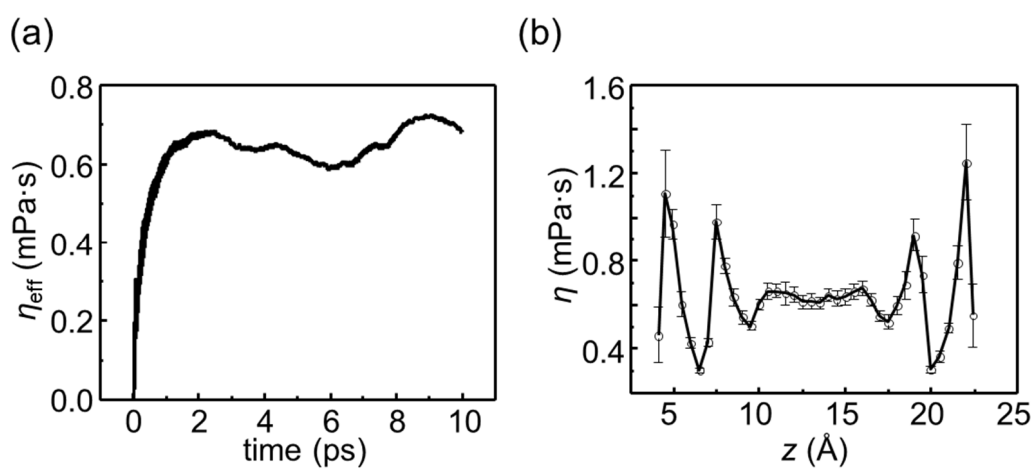

**Figure S8. The quantitative local viscosity distribution for water flow of the system water sheared by graphene. a,** The viscosity for the whole water molecules. **b,** The quantitative local viscosity distribution for water.

For water on FDTS monolayer sheared by silica, the quantitative local viscosity distribution was calculated with the standard process, as shown in Fig. S9.

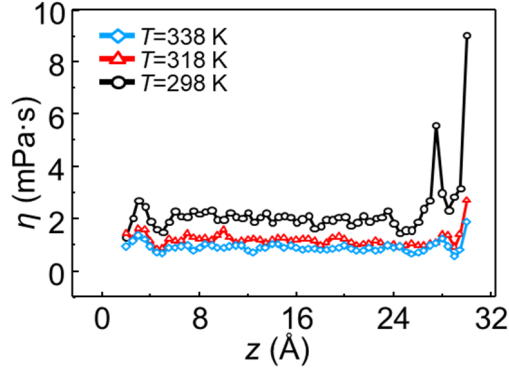

**Figure S9. The quantitative local viscosity distribution for water flow of the system water on FDTS monolayer sheared by silica.**

### 7. Slip length calculations

In this section, we give the details of two methods for calculating slip length $l_s$ used in Fig. 3g of the main text. One is to obtain $l_s$ from the given velocity profile, the other is to obtain $l_s$ with equilibrium molecular dynamics simulation.

**a.** The method with velocity profile

The method in Ref. (*16*) for calculating the slip length $l_s$ was used here, $l_s = v_{\text{slip}}/\dot{\gamma}$, where $v_{\text{slip}}$ is the slip velocity and $\dot{\gamma}$ is the shear rate. $v_{\text{slip}}$ was obtained as the difference between $V_{\text{wall}}$ and the extrapolated bulk velocity at the hydrodynamic wall position (HWP) $V_{\text{HWP}}$, i.e., $v_{\text{slip}}=V_{\text{wall}}-V_{\text{HWP}}$. $V_{\text{HWP}} = \dot{\gamma}h/2$, where $h = N/n_{\text{bulk}}S$, $N$ is the number of molecules, $n_{\text{bulk}}$ is the number density of bulk fluid, and $S$ is the area of the wall. For velocity profile obtained from both APM and conventional method, the same calculation process is used to obtain $l_s$.

**b.** The method with equilibrium molecular dynamics (EMD)

The EMD method calculates the slip length as $l_s = \eta_{bulk}/\lambda$, where $\eta_{\text{bulk}}$ is the viscosity of the bulk fluid and $k$ is the friction coefficient. We refer to Ref. (*17*) for calculating the friction coefficient as

$$k = \lim_{t \to \infty} \frac{1}{Sk_BT} \int_0^t \sum_i \langle F_i(0)F_i(\tau) \rangle_{\text{EC}} d\tau, \qquad (24)$$

where $F_i$ is the force between the single molecule $i$ and the wall, $\langle \rangle_{\text{EC}}$ denotes the ensemble average at the equilibrium condition, and $S$ is the area of the wall. Equilibrium molecular dynamics simulation was conducted with the same procedure of simulations in the Supplementary Section 4 except for the shearing. The data collecting lasted for 20 ps with a time step of 1 fs.

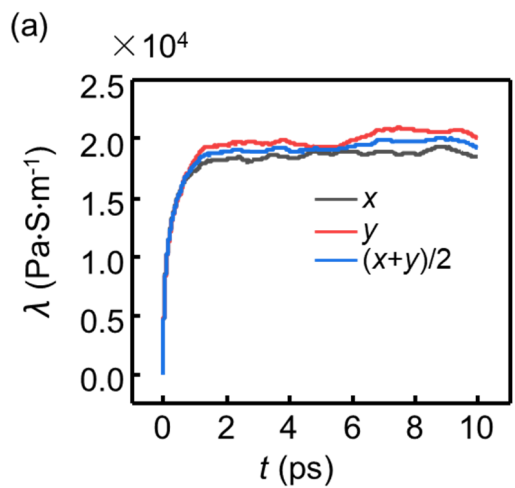

**Figure S10. Friction coefficient from the equilibrium molecular dynamics method.**

For water confined by the graphene, the results of friction coefficient were shown in Fig. S10. As the upper limit of integral $t$ could not be infinite in practice, here we plotted $k$ as a function of $t$ to check the convergence of equation (14). The force autocorrelation function (FAF) converges at ~2 ps.

## 8. Computing time estimation

In this section, we give the details for the estimate of the computing time to obtain the velocity profile shown in Fig. 3b of the main text with the conventional method.

The ratio of standard error of APM and the conventional method $\gamma$ is $6.1 \times 10^{-7}$ for $V_{wall} = 10^{-4}$ m/s. The standard error is inverse proportional to the square root of the number of independent samples. Therefore, the number of samples for conventional method to obtain the same standard error would be $1/\gamma^2 = 2.7 \times 10^{12}$ times the number of samples for APM which was generated by a simulation of 1 ns. This corresponds to a simulation of $10^{-9} \times 2.7 \times 10^{12} = 2.7 \times 10^3$ s. In practical MD simulation of present setup, 7.5 hours is needed to conduct a simulation of 1 ns on a node with 32 cores running at 2.9 GHz. This amounts to $7.4 \times 10^{10}$ CPU-years of simulation time in total, where one

CPU-year is equivalent to one 2.9 GHz CPU core running continuously for a year.

## 9. The experimental measuring details

In this section, we firstly give the experimental measuring details in the Fig. 4a of the main text. Then we give the relation between the normal velocity and shearing velocity of the FDTS monolayer surface.

**a.** Experimental setup

The friction coefficient was measured by the atomic force microscope (AFM, Oxford Instrument, Cypher ES). The microsphere (glass, radius 10 μm) was glued on a tipless cantilever (CSC38, Mikromasch Instrument) of AFM. The FDTS monolayer was physically vapor-deposited onto a silica substrate. The water applied for friction coefficient measurement was degassed deionized water (18.2 MΩ, Hitech Sciencetool). During the measurement, the microsphere and the FDTS monolayer were immersed into water. The FDTS monolayer was driven by the piezoelectric ceramic scanner to approach the microsphere. Meanwhile, the force acting on microsphere ($F$) and the separation ($D$) between microsphere and FDTS monolayer were recorded by the optical lever system of AFM. A typical force-distance curve ($F$-$D$) was shown in Fig. S11. For such sphere-plane squeeze-out setup, $V_n/F = (D + l_s)/6\pi\eta R^2$, where $V_n = dD/dt$ is the normal velocity of FDTS monolayer to approach the microsphere, $t$ is time, $l_s$ is the slip length at water-FDTS interface, $R$ is the radius of microsphere, $\eta$ is the viscosity of water. Thus, the friction coefficient $\lambda = \eta/l_s$ can be calculated by the vertical intercept $b$ of $V_n/F$-$D$ curves as $\lambda = 1/6\pi R^2 b$.

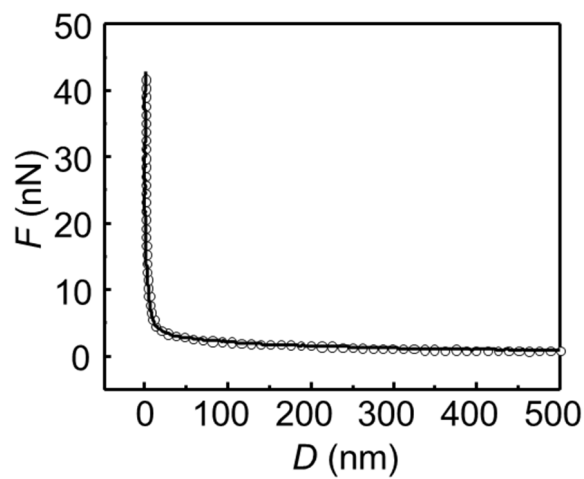

**Figure S11. The force-distance curves in experiments.**

**b.** The relation between normal velocity and shearing velocity

The derivation here was borrowed from Ref. (*18*). The flow field is formed by the squeezing of the microsphere and the FDTS monolayer substrate in experiments. Technically, the water flow field can be described by the model as shown in Fig. S12. Following are the definitions of cylindrical system (*z*, *r*) of coordinates: (1) the axis *z* coincides with the line connecting the centers of spheres and perpendicular to the substrate. (2) the axis *r* is parallel to the substrate. (3) the plane *z* = 0 is the surface of the substrate.

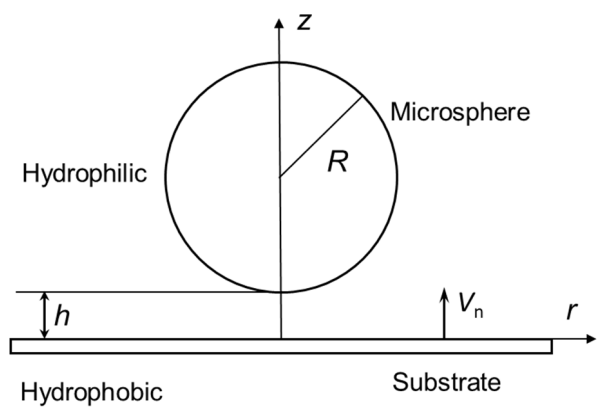

**Figure S12. The theoretical model for calculating water flow field in the experiments.**

The velocity of water at the *z* = 0 surface can be expressed as (*18*):

$$v_r = \frac{1}{2\mu}\frac{\partial p}{\partial r}\left(\frac{l_s H(H + 2l_s(1+k))}{H + l_s(2+k)}\right), \tag{25}$$

where $\mu$ is the viscosity, $p$ is the pressure of water, $l_s$ is the slip length of substrate, $H = h + \frac{1}{2}\frac{r^2}{R}$ and $k$ is the parameter that reflects the slip length of the microsphere. For the hydrophilic surface of microsphere, $k = -1$.

$$v_r = \frac{1}{2\mu}\frac{\partial p}{\partial r}\left(\frac{l_s H^2}{H + l_s}\right). \tag{26}$$

The pressure $p = -\frac{3\mu R V_n}{H^2}p^*$ (*18*), where $p^*$ is the dimensionless number. The value of $p^*$ can be obtained by looking up the table in Ref. (*18*), i.e., $p^* = 0.5$.

$$v_r = \frac{3V_n r l_s}{2H(H+l_s)}. \tag{27}$$

$v_r$ changes with the value of $r$, here we chose a fixed value of $r = 0.45R$.

$$v_s = \frac{l_s + h}{l_s} v_r. \tag{28}$$

Here, $v_s$ is the corresponding shearing velocity in the Couette model as shown in Fig. S2b. For the typical value in this work, $l_s = 50$ nm, $R = 10$ μm and $h = 50$ nm,

$$v_s \approx 0.6 V_n. \tag{29}$$

## 10. Description of supporting movies

Movies S1: The flow fields of water sheared by graphene with $V_{wall} = 10^{-4}$ m/s for APM.

Movies S2: The flow fields of water sheared by graphene with $V_{wall} = 10^{-3}$ m/s for APM.

Movies S3: The flow fields of water sheared by graphene with $V_{wall} = 100$ m/s for APM.

Movies S4: The flow fields of water sheared by graphene with $V_{wall} = 100$ m/s for CON.